\documentclass[11pt]{article}
\usepackage{graphicx}
\usepackage{amsmath, amsfonts, amssymb, amsthm}
\usepackage[english]{babel}
\usepackage{enumerate}
\usepackage{booktabs}
\usepackage{caption}
\usepackage{subcaption}
\usepackage{epstopdf}
\usepackage{natbib}
\usepackage{multicol}
\usepackage{etoolbox}
\usepackage{relsize}

\newcommand{\blind}{0}

\numberwithin{equation}{section}
\theoremstyle{plain}
\newtheorem{theorem}{Theorem}[section]
\newtheorem{lemma}[theorem]{Lemma}
\newtheorem{prop}[theorem]{Proposition}

\theoremstyle{definition}
\newtheorem{example}{Example}[section]
\newtheorem{defn}{Definition}[section]

\newenvironment{remark}[1][Remark.]{\begin{trivlist}
\item[\hskip \labelsep {\bfseries #1}]}{\end{trivlist}}

\newcommand{\bs}[1]{\boldsymbol{#1}}

\newcommand{\te}[1]{\textnormal{#1}}
\newcommand{\nn}{\nonumber \\}
\newcommand{\tr}{\textnormal{Tr}}

\newcommand{\Log}{\textnormal{Log}}
\newcommand{\Exp}{\textnormal{Exp}}

\makeatletter
\newcommand*\rel@kern[1]{\kern#1\dimexpr\macc@kerna}
\newcommand*\widebar[1]{%
  \begingroup
  \def\mathaccent##1##2{%
    \rel@kern{0.8}%
    \overline{\rel@kern{-0.8}\macc@nucleus\rel@kern{0.2}}%
    \rel@kern{-0.2}%
  }%
  \macc@depth\@ne
  \let\math@bgroup\@empty \let\math@egroup\macc@set@skewchar
  \mathsurround\z@ \frozen@everymath{\mathgroup\macc@group\relax}%
  \macc@set@skewchar\relax
  \let\mathaccentV\macc@nested@a
  \macc@nested@a\relax111{#1}%
  \endgroup
}
\makeatother
\begin{document}

\bibliographystyle{chicago}

\def\spacingset#1{\renewcommand{\baselinestretch}%
{#1}\small\normalsize} \spacingset{1}


\if0\blind
{
  \title{\bf  Intrinsic data depth for Hermitian positive definite matrices}
  \author{Joris Chau\footnote{
	Corresponding author, j.chau@uclouvain.be, Institute of Statistics, Biostatistics, and Actuarial Sciences, Universit\'e catholique de Louvain, Voie du Roman Pays 20, B-1348, Louvain-la-Neuve, Belgium. 
  },\quad Hernando Ombao\footnote{
  Department of Statistics, University of California at Irvine, Bren Hall 2206, Irvine, CA, 92697, United States.
  Department of Applied Mathematics and Computational Science, King Abdullah University of Science and Technology, Thuwal 23955-6900, Saudi Arabia.
    }\quad and\ Rainer von Sachs\footnote{
  Institute of Statistics, Biostatistics, and Actuarial Sciences, Universit\'e catholique de Louvain, Voie du Roman Pays 20, B-1348, Louvain-la-Neuve, Belgium.  
    }
  }
  \date{}
  \maketitle
} \fi

\if1\blind
{
  \bigskip
  \bigskip
  \bigskip
  \begin{center}
    {\LARGE\bf Intrinsic data depth for Hermitian positive definite matrices}
\end{center}
  \medskip
} \fi

\bigskip

\begin{abstract} 
\noindent 
Nondegenerate covariance, correlation and spectral density matrices are necessarily symmetric or Hermitian and positive definite. The main contribution of this paper is the development of statistical data depths for collections of Hermitian positive definite matrices by exploiting the geometric structure of the space as a Riemannian manifold. The depth functions allow one to naturally characterize most central or outlying matrices, but also provide a practical framework for inference in the context of samples of positive definite matrices. First, the desired properties of an intrinsic data depth function acting on the space of Hermitian positive definite matrices are presented. Second, we propose two computationally fast pointwise and integrated data depth functions that satisfy each of these requirements and investigate several robustness and efficiency aspects. As an application, we construct depth-based confidence regions for the intrinsic mean of a sample of positive definite matrices, which is applied to the exploratory analysis of a collection of covariance matrices associated to a multicenter research trial. 
\end{abstract}

\noindent%

\spacingset{1.45} 

\section{Introduction} \label{sec:1}
In numerous applications in multivariate statistics, we are interested not only in the first-order behavior (mean) of a sample of random vectors, but also in the second-order behavior or variability of the sample. In fact, our primary interest is often precisely the analysis of covariance or correlation structures between components of the random vectors. In many areas of statistical research, such as neuroscience, biomedical science, environmental science, demographics or finance, it is increasingly common to encounter covariance or correlation matrices across a large number of temporal or spatial locations, or across a large number of replicated subjects or trials in an experiment. In this work, our aim is to develop data exploration and inference tools for large collections or samples of such matrices.\\[3mm]
The data objects of interest, nondegenerate covariance or correlation matrices, are necessarily elements of the space of Hermitian positive definite (HPD) matrices, which includes the space of symmetric positive definite (SPD) matrices in the real-valued case. The space of HPD matrices, although very well-structured, is inherently non-Euclidean and standard Euclidean-based statistical procedures (e.g., regression, clustering or inference procedures) may be unstable or break down due to the geometric constraints of the space. For this reason, it is necessary to generalize statistical procedures for data in the space of symmetric or Hermitian PD matrices, taking into account the non-Euclidean geometry of the space. Several recent works addressing this issue include: \cite{S00}, \cite{PFA05}, \cite{F09}, \cite{Z09}, \cite{D09}, \cite{F11}, \cite{Y12} \cite{S15}, \cite{H16} and \cite{CvS17} among others. The main contribution of this paper is the generalization of notions of data depth for samples of HPD matrices to provide a center-to-outward ordering of positive definite matrix-valued objects. \\[3mm]
Data depth is a useful tool for data exploration to identify most central or outlying data observations (as in \cite{L99} or \cite{SG12} in a Euclidean context); or as a means of inference, by way of rank-based hypothesis testing (as in \cite{LS93}, \cite{CS12}, or \cite[Chapter 5]{M02}), classification (see e.g., \cite{L12}) or the construction of confidence regions (see e.g., \cite{YS97}) among other applications. Although many different depth functions have been proposed and studied in the literature over the years, most data depth functions are constructed in the first place for vector-valued observations in the Euclidean space $\mathbb{R}^d$. Exceptions include \cite{LS92}, where the authors consider depth functions for directional data on circles or spheres; \cite{H11}, on projection depth for tensor objects; or the recent work, \cite{PB17}, on halfspace depths for scatter, concentration and shape matrices. For an overview of various Euclidean data depth functions and their specific properties, we refer the reader to e.g., \cite{L99}, \cite{ZS00}, or \cite{M02}. \\[3mm]
The space of $(d \times d)$-dimensional Hermitian (not necessarily PD) matrices $(\mathbb{H}_{d \times d}, +, \cdot_S)$ together with matrix addition and matrix scalar multiplication is a real vector space, and each Hermitian matrix bijectively maps to a vector in $\mathbb{R}^{d^2}$ by expanding the matrix with respect to some basis. To calculate data depth values for a sample of Hermitian matrices, it suffices to apply any ordinary Euclidean data depth function to the basis component vectors of the Hermitian matrices, given that the computed depth values do not depend on the chosen basis. In contrast, due to the nonlinear positive definite constraints, the space of HPD matrices $(\mathbb{P}_{d \times d}, +, \cdot_S)$ is not a vector space. Moreover, the cone of HPD matrices embedded in a Euclidean space endowed with the Eulidean metric is not a complete metric space. As a consequence, Euclidean data depth applied to a sample of HPD matrices violates the basic properties of a proper depth function. To illustrate, according to \cite{ZS00}, a proper depth function should be monotonicallly non-increasing moving outwards from a well-defined center. Moving away from a central point along a straight line is not always well-defined in the cone of HPD matrices, as the boundary of the space lies at a finite distance. Also, pointwise or uniform continuity properties of the data depth functions fail to hold due to the incompleteness of the metric space. \\[3mm]
Instead of embedding the space of HPD matrices in an ambient Euclidean space, we exploit the geometric structure of the space of HPD matrices as a curved Riemannian manifold equipped with the affine-invariant (\cite{PFA05}) --also natural invariant (\cite{S00}), canonical (\cite{H16}), trace (\cite{Y12}), Rao-Fisher (\cite{S15})-- Riemannian metric, or simply the Riemannian metric (\cite{B09}, \cite{D09}). The affine-invariant metric plays an important role in estimation problems in the space of symmetric or Hermitian PD matrices for several reasons: (i) the space of HPD matrices equipped with the affine-invariant metric is a complete metric space, (ii) the affine-invariant metric is invariant under congruence transformation by any invertible matrix, see Section \ref{sec:2}, and (iii) there is no \emph{swelling effect} as with the Euclidean metric, where interpolating two HPD matrices may yield a matrix with a determinant larger than either of the original matrices, which may lead to computational instability, (\cite{P10}). The first property allows us to construct proper data depth functions in the space of HPD matrices satisfying all of the intrinsic versions of the axiomatic properties in \cite{ZS00}. The second property is important to ensure that the depth functions are general linear congruence invariant, which in practice means that the depth values do not non-trivially depend on the chosen coordinate system of the data. In \cite{D09}, the authors list several additional metrics for estimation problems in the space of HPD matrices, such as the Log-Euclidean metric, also studied in \cite{A06}. The Log-Euclidean metric transforms the space of HPD matrices into a complete metric space and is invariant under congruence transformations by the unitary group, but not by the general linear group, as is true for the affine-invariant metric.\\[3mm]
In the preliminary Section \ref{sec:2}, we introduce the necessary geometric tools to develop data depths acting directly on the space of HPD matrices as a geodesically complete manifold. In Section \ref{sec:3}, we present the desired properties an intrinsic depth function should satisfy, and we propose two data depth functions that satisfy each of these requirements. In addition, we consider integrated depth functions that act on curves of HPD matrices, such as spectral density matrices. In Section \ref{sec:4}, we compare the two depth functions in terms of robustness and efficiency aspects. In Section \ref{sec:5}, as an application of the depth functions, we construct depth-based confidence regions for the intrinsic mean of a sample of HPD matrices, and in Section \ref{sec:6} we apply the intrinsic depth functions to explore a collection of covariance matrices from a multicenter clinical trial. The technical proofs and additional figures can be found in the supplementary material. The accompanying \texttt{R}-code, containing the necessary tools to compute the intrinsic data depths and to perform rank-based hypothesis testing for samples of HPD matrices, is publicly available in the \texttt{R}-package \texttt{pdSpecEst} on CRAN, (\cite{C17}).
\section{Preliminaries} \label{sec:2}
\subsection{Geometry of HPD matrices}
In order to develop data depths for observations in the space of HPD matrices, we study the space as a Riemannian manifold as in \cite{PFA05}, \cite[Chapter 6]{B09}, or \cite{S00} among others. Denote $\mathcal{M} := \mathbb{P}_{d \times d}$ for the space of $(d \times d)$ HPD matrices. $\mathcal{M}$ is an open subset of the space $(d \times d)$ Hermitian matrices $\mathcal{H} := \mathbb{H}_{d \times d}$, and as such a smooth manifold. The tangent space $T_p(\mathcal{M})$ at a point (i.e., a matrix) $p \in \mathcal{M}$ can be identified by the Hermitian space $\mathcal{H}$, and the Frobenius inner product on $\mathcal{H}$ induces the affine-invariant Riemannian metric $g_R$ on the manifold $\mathcal{M}$ given by the smooth family of inner products:
\begin{eqnarray} \label{eq:2.1}
\langle h_1, h_2 \rangle_p &=& \tr((p^{-1/2} \ast h_1)(p^{-1/2} \ast h_2)), \quad \quad \forall\: p \in \mathcal{M},
\end{eqnarray}
with $h_1,h_2 \in T_p(\mathcal{M})$. Here and throughout this paper, $y^{1/2}$ always denotes the Hermitian square root matrix of $y \in \mathcal{M}$, and we write $y \ast x := y^* x y$ for matrix congruence transformation, where $^*$ denotes the conjugate transpose of a matrix. The Riemannian distance $\delta_R$ on $\mathcal{M}$ derived from the affine-invariant Riemannian metric is given by:
\begin{eqnarray}
\delta_R(p_1,p_2) &=& \Vert \Log(p_1^{-1/2} \ast p_2) \Vert_F,  \label{eq:2.2}
\end{eqnarray}
where $\Vert \cdot \Vert_F$ denotes the matrix Frobenius norm and $\Log(\cdot)$ is the matrix logarithm. Denote the general linear group by $\te{GL}(d, \mathbb{C}) := \{ a \in \mathbb{C}^{d \times d}\, :\, \det(a) \neq 0 \}$. The mapping $x \mapsto a \ast x$ is an isometry for each invertible matrix $a \in \te{GL}(d,\mathbb{C})$, i.e., it is distance-preserving:
\begin{eqnarray*}
\delta_R(p_1, p_2) &=& \delta_R(a \ast p_1, a \ast p_2),\quad \forall\: a \in \te{GL}(d, \mathbb{C}).
\end{eqnarray*}
By \cite[Theorem 6.1.6 and Prop. 6.2.2]{B09}, the Riemannian manifold $(\mathcal{M}, g_R)$ is geodesically complete. By the Hopf-Rinow Theorem this implies that there exists a unique \emph{geodesic} segment joining any two points $p_1,p_2 \in \mathcal{M}$ and every geodesic can be extended indefinitely. The Hopf-Rinow Theorem also implies that for every $p \in \mathcal{M}$ the exponential map $\Exp_p$ and the logarithmic (i.e., inverse exponential) map $\Log_p$ are global diffeomorphisms with domains $T_p(\mathcal{M})$ and $\mathcal{M}$ respectively. By (\cite{PFA05}), the exponential $\Exp_p: T_p(\mathcal{M}) \to \mathcal{M}$ and logarithmic $\Log_p : \mathcal{M} \to T_p(\mathcal{M})$ maps are given by,
\begin{eqnarray*}
\Exp_p(h) &=& p^{1/2} \ast \Exp\left(p^{-1/2} \ast h \right),\nn
\Log_p(q) &=& p^{1/2} \ast \Log\left(p^{-1/2} \ast q \right),
\end{eqnarray*}
where $\Exp(\cdot)$ denotes the matrix exponential. The Riemannian distance may now also be expressed in terms of the logarithmic map as:
\begin{eqnarray} \label{eq:2.3}
\delta_R(p_1, p_2) \ \ = \ \ \Vert \Log_{p_1}(p_2) \Vert_{p_1} \ \ = \ \ \Vert \Log_{p_2}(p_1) \Vert_{p_2}, \quad \forall\: p_1, p_2 \in \mathcal{M},
\end{eqnarray}
where throughout this paper $\Vert h \Vert_p := \langle h, h \rangle_p$ denotes the norm of $h \in T_p(\mathcal{M})$ induced by the affine-invariant metric.\\[3mm]
As there exists a unique geodesic curve connecting any two points $p_1, p_2 \in \mathcal{M}$, geodesically convex sets are well-defined. A subset $\mathcal{K} \subseteq \mathcal{M}$ is said to be convex or geodesically convex if for each pair of points $p_1,p_2 \in \mathcal{K}$, the geodesic segment $[p_1, p_2]$ is contained entirely in $\mathcal{K}$. If $\mathcal{S} \subseteq \mathcal{M}$, then the convex hull of $\mathcal{S}$, denoted by $\te{conv}(\mathcal{S})$, is the smallest convex set containing $\mathcal{S}$. This set is conveniently expressed as,
\begin{eqnarray*}
\te{conv}(\mathcal{S}) := \left\{ p \in \mathcal{S}: p = \Exp_p\left(\int_{\mathcal{S}} \Log_p(x) w(x)\ \lambda(dx) \right),\ w: \mathcal{S} \to [0,1],\ \int_{\mathcal{S}} w(x)\ \lambda(dx) = 1 \right\}, 
\end{eqnarray*} 
where $\lambda$ is the Lebesgue measure on the finite-dimensional metric space $(\mathcal{M}, \delta_R)$ and $w$ is a measurable function. For more details on the construction of (approximate) convex hulls on the manifold $\mathcal{M}$, we refer to \cite{F11}.

\subsection{Probability distributions and random variables}
A random variable $X: \Omega \to \mathcal{M}$ on the Riemannian manifold $(\mathcal{M}, g_R)$ is a measurable function from some probability space $(\Omega, \mathcal{A}, \nu)$ to the measurable space $(\mathcal{M}, \mathcal{B}(\mathcal{M}))$, where $\mathcal{B}(\mathcal{M})$ is the Borel algebra, i.e., the smallest $\sigma$-algebra containing all open sets in $(\mathcal{M}, g_R)$. In the following, we always work directly with the induced probability on $\mathcal{M}$, $\nu(B) = \nu(\{ \omega \in \Omega: X(\omega) \in B \})$. By $P(\mathcal{M})$, we denote the set of all probability measures on $(\mathcal{M}, \mathcal{B}(\mathcal{M}))$ and $P_p(\mathcal{M})$ denotes the subset of probability measures in $P(\mathcal{M})$ that have finite moments of order $p$ with respect to the Riemannian distance, i.e., the $L^p$-Wasserstein space \cite[Definition 6.4]{V09}:
\begin{eqnarray*}
P_p(\mathcal{M}) &:=& \left\{ \nu \in P(\mathcal{M}) : \exists\: y_0 \in \mathcal{M},\ \te{s.t.} \int_{\mathcal{M}} \delta_R(y_0, x)^p\: \nu(dx) < \infty \right\}.
\end{eqnarray*}
Note that if $\int_{\mathcal{M}} \delta_R(y_0, x)^p\: \nu(dx) < \infty$ for some $y_0 \in \mathcal{M}$ and $1 \leq p < \infty$, this is true for any $y \in \mathcal{M}$. This follows by the triangle inequality and the fact that $\delta_R(p_1, p_2) < \infty$ for any $p_1,p_2 \in \mathcal{M}$, as $\int_{\mathcal{M}} \delta_R(y, x)^p\: \nu(dx) \leq 2^p \left( \delta_R(y, y_0)^p + \int_{\mathcal{M}} \delta_R(y_0, x)^p \: \nu(dx) \right) < \infty$. For a sequence of probability measures $(\nu_n)_{n \in \mathbb{N}}$ in $P(\mathcal{M})$, $\nu_n \overset{w}{\to} \nu$ denotes weak convergence to the probability measure $\nu$ in the usual sense, i.e., $\int_{\mathcal{M}} \phi(x)\: \nu_n(dx) \to \int_{\mathcal{M}} \phi(x)\: \nu(dx)$ for every continuous and bounded function $\phi: \mathcal{M} \to \mathbb{R}$, and a sequence $(\nu_n)_{n \in \mathbb{N}}$ is said to be uniformly integrable if $\lim_{K \to \infty} \sup_{n \in \mathbb{N}} \int_{\mathcal{M}} \delta_R(y_0, x) \bs{1}_{\{\delta_R(y_0, x) > K \}}\: \nu_n(dx) = 0$ for some $y_0 \in \mathcal{M}$. Note that if $(\nu_n)_{n \in \mathbb{N}}$ is uniformly integrable for some $y_0 \in \mathcal{M}$, then the sequence is uniformly integrable for any $y \in \mathcal{M}$. Finally, we use the notation $\te{conv}(\nu) := \te{conv}(\te{supp}(\nu))$ for the convex hull of the support of the measure $\nu$ on $\mathcal{M}$, and $\te{rint}(\te{conv}(\nu))$ and $\te{r}\delta(\te{conv}(\nu))$ for its relative interior and relative boundary.

\subsection{Measures of centrality}
\paragraph{Intrinsic mean.}
To characterize the center of a random variable $X$ with probability measure $\nu$, one important measure of centrality is the Karcher or Fr\'echet mean, which is also referred to as the intrinsic mean as it is \emph{intrinsic} to the Riemannian distance measure on the manifold. The intrinsic mean turns out to be the point of maximum depth in the intrinsic zonoid depth introduced in Section \ref{sec:3.2}. The set of intrinsic means consists of the points that minimize the second moment with respect to the Riemannian distance, 
\begin{eqnarray*}
\mu\ =\ \mathbb{E}_{\nu}[X] \ := \ \arg\min_{y \in \te{supp}(\nu)} \int_\mathcal{M} \delta_R(y,x)^2\ \nu(dx).
\end{eqnarray*}
If $\nu \in P_2(\mathcal{M})$, then at least one intrinsic mean exists as the above expectation is finite for $y \in \mathcal{M}$. Moreover, since the manifold $\mathcal{M}$ is a geodesically complete manifold of non-positive curvature (see \cite{PFA05} or \cite{S84}), by \cite[Proposition 1]{Le95} the intrinsic mean $\mu$ is unique for any distribution $\nu \in P_2(\mathcal{M})$. By \cite[Corollary 1]{P06}, the intrinsic mean is also represented by the point $\mu \in \mathcal{M}$ that satisfies,
\begin{eqnarray} \label{eq:2.4}
\bs{E}_\nu[\Log_{\mu}(X)] &=& \bs{0},
\end{eqnarray}
where $\bs{0}$ is the zero matrix. The sample intrinsic mean of a set of manifold-valued observations minimizes a sum of squared Riemannian distances and can be computed efficiently through a gradient descent algorithm as in \cite{P06}.
\paragraph{Intrinsic median.}
A second measure of centrality of primary interest is the intrinsic median as in \cite{F09}, which is the point of maximum depth in the geodesic distance depth defined in Section \ref{sec:3.4}. The set of intrinsic medians minimizes the first moment with respect to the Riemannian distance,
\begin{eqnarray*}
m \ = \ \te{GM}_{\nu}(X) \ :=\ \arg\min_{y \in \te{supp}(\nu)} \int_\mathcal{M} \delta_R(y, x)\: d\nu(x).
\end{eqnarray*}
On $(\mathcal{M}, \delta_R)$, a geodesically complete manifold with non-positive curvature, the intrinsic median exists and is unique for any distribution $\nu \in P_1(\mathcal{M})$. This follows by the proof of \cite[Theorem 1]{F09} combined with an application of Leibniz's integral rule. Furthermore, the intrinsic median is uniquely characterized by the point $m \in \mathcal{M}$ that satisfies, 
\begin{eqnarray} \label{eq:2.5}
\bs{E}_{\nu}\left[ \frac{\Log_m(X)}{\delta_R(m, X)} \right] &=& \bs{0}.
\end{eqnarray}
\begin{remark}
If the distribution $\nu$ of a random variable $X$ is \emph{centrally symmetric} around $\mu \in \mathcal{M}$ in the sense that $\Log_\mu(X) \overset{d}{=} - \Log_{\mu}(X)$, then the intrinsic mean and median coincide and are equal to $\mu$. Here, equality in distribution ($\overset{d}{=}$) is read as equality in terms of the joint distribution of all matrix components. The claim for the intrinsic mean follows by the fact that $\bs{E}_\nu[\Log_{\mu}(X)] = \bs{0}$, which implies that $\mu$ is the intrinsic mean of the random variable $X$. For the intrinsic median, if $X$ is centrally symmetric around $\mu$, then $X$ is also \emph{angularly symmetric} around $\mu$ in the sense that $\Log_{\mu}(X)/\Vert \Log_\mu(X) \Vert_\mu \overset{d}{=} - \Log_{\mu}(X)/\Vert \Log_\mu(X) \Vert_\mu$. Substituting $\Vert \Log_\mu(X) \Vert_\mu = \delta_R(\mu, X)$, we observe that $\bs{E}_\nu[\Log_{\mu}(X)/\delta_R(\mu, X)] = \bs{0}$, which implies that $\mu$ is also the intrinsic median of the random variable $X$. 
\end{remark}
\section{Data depth for HPD matrices} \label{sec:3}
Before introducing the manifold data depth functions, we present the desired properties a proper intrinsic data depth function --acting directly on the space of HPD matrices-- should satisfy. These requirements are the natural intrinsic generalizations of the properties in \cite{ZS00} for depth functions acting on vectors in a Euclidean space $\mathbb{R}^d$. We also consider integrated analogs for depth functions acting on curves of HPD matrices $y(t) \in \mathcal{M}$ with $t \in \mathcal{I} \subset \mathbb{R}$, such as spectral density matrices in the Fourier domain.
\subsection{Depth properties} \label{sec:3.1}
Below, we denote $\te{D}(\nu, y)$ for the depth of a matrix $y \in \mathcal{M}$ with respect to a distribution $\nu \in P(\mathcal{M})$; or $\te{iD}(\nu, y)$ for the integrated depth of a matrix curve $y := (y(t))_{t \in \mathcal{I}}$ with respect to a curve of marginal measures $\nu := (\nu(t))_{t \in \mathcal{I}}$, such that $\nu(t) \in P(\mathcal{M})$ for each $t \in \mathcal{I}$. If a nonnegative bounded function $\te{D}(\cdot, \cdot)$ or $\te{iD}(\cdot, \cdot)$ satisfies the pointwise (resp.\@ integrated) properties \textbf{P.1} to \textbf{P.4}, we say that it is a proper data depth function on the Riemannian manifold $(\mathcal{M}, g_R)$.\\[3mm]
\textbf{P.1} \emph{(Congruence invariance)} The depth function should be invariant under matrix congruence transformation of the form $x \mapsto a \ast x$, with $a \in \te{GL}(d, \mathbb{C})$. That is, for each $a \in \te{GL}(d, \mathbb{C})$,
\begin{eqnarray} \label{eq:3.1}
\te{D}(\nu, y) &=& \te{D}(\nu_a, a \ast y), \quad \forall\: y \in \mathcal{M},
\end{eqnarray} 
where $\nu_a$ is the distribution of the transformed random variable $a \ast X$, such that $X$ is distributed according to $\nu$. Generalizing this property for an integrated depth function $\te{iD}(\nu, y)$, we require that the same property holds pointwise for each $t \in \mathcal{I}$. In this case, $a := (a(t))_{t \in \mathcal{I}}$ is a curve of invertible matrices, with $a(t) \in \te{GL}(d, \mathbb{C})$ for each $t \in \mathcal{I}$.\\[3mm]
In a standard Euclidean context, for a depth function acting on vectors in the Euclidean space $\mathbb{R}^d$, it is desirable that the depth is affine-invariant $\te{D}(\nu, y) = \te{D}(\nu_{a,b}, ay + b)$ for each $y \in \mathbb{R}^d$, where $\nu_{a,b}$ is the distribution of the random vector $aX + b$, with $a \in \te{GL}(d, \mathbb{R})$, $b \in \mathbb{R}^d$ and $X$ distributed according to $\nu$. In the current setup, we are concerned with covariance or correlation matrices, corresponding to the second-order behavior of a random vector. For a random vector $X$ with covariance matrix $\Sigma$, the covariance matrix of the affine transformation $aX + b$ is given by $a^T \ast \Sigma = a \Sigma a^T$. A natural requirement for the depth functions acting on symmetric or Hermitian PD matrices is therefore invariance under congruence transformations of the data. Another way to view this is that a depth function acting on the covariance matrix of a data vector $X$ should be invariant under a change of basis in the data space of $X$. \\[3mm]
\textbf{P.2} \emph{(Maximality at center)} The depth function should attain its maximum value, i.e., deepest point, at a well-defined unique \emph{center} of the distribution, such as the intrinsic mean or median, which are characterized as the points of central and angular symmetry respectively. Let $\mu \in \mathcal{M}$ be a unique central point of the distribution $\nu$, then,
\begin{eqnarray*}
\te{D}(\nu, \mu) &=& \sup_{y \in \mathcal{M}} \te{D}(\nu, y).
\end{eqnarray*}
Similarly, for an integrated depth function, the maximum value should be attained at a well-defined  unique central curve $\mu(t)$ with $t \in \mathcal{I}$, such as the curve of intrinsic means or medians.\\[3mm]
\textbf{P.3} \emph{(Monotonicity relative to center)} As $y \in \mathcal{M}$ moves away from the deepest point $\mu$ along a geodesic curve emanating from $\mu$, the depth of the point $y$ with respect to the distribution $\nu$ should be monotonically non-increasing. Let $\Exp_\mu(th)$, $t \geq 0$, be the geodesic emanating from $\mu$ with unit tangent vector $h$. Then,
\begin{eqnarray*}
\te{D}(\nu, \Exp_{\mu}(t_1 h)) & \geq & \te{D}(\nu, \Exp_{\mu}(t_2 h)), \quad \forall\: 0 \leq t_1 \leq t_2.
\end{eqnarray*}
For an integrated depth function, let $s_1(t), s_2(t)$ be real-valued curves over $\mathcal{I}$, such that $0 \leq s_1(t) \leq s_2(t)$ for each $t \in \mathcal{I}$. Denote $y_1(t) := \Exp_{\mu(t)}(s_1(t) h(t))$ and $y_2(t) := \Exp_{\mu(t)}(s_2(t) h(t))$, where $h(t) \in T_{\mu(t)}(\mathcal{M})$ is a curve of unit tangent vectors. Then, 
\begin{eqnarray*}
\te{iD}(\nu, y_1) & \geq & \te{iD}(\nu, y_2).
\end{eqnarray*}
\textbf{P.4} \emph{(Vanishing at infinity)} The depth of a point $y \in \mathcal{M}$ should approach zero as the point $y$ converges to a singular matrix, i.e., a matrix with zero or infinite eigenvalues, 
\begin{eqnarray*}
\lim_{M \to \infty} \sup_{\Vert \Log(y) \Vert_F \geq M}  \te{D}(\nu, y) &=& 0.
\end{eqnarray*}
Similarly, for an integrated depth function, if the curve $y(t)$ converges to a curve of singular matrices for each $t \in \mathcal{I}$, then the integrated depth should approach zero. \\[3mm]
Below, we give two additional continuity properties, which although not strictly required are nonetheless useful to derive asymptotic results in subsequent applications, such as rank-based hypothesis testing or the construction of depth-based confidence sets as in Section \ref{sec:5}. \\[3mm]
\textbf{(P.5)} \emph{(Continuity in $y$)} Let $(y_n)_{n \in \mathbb{N}}$ be a convergent sequence with $y_n \in \mathcal{M}$ for each $n \in \mathbb{N}$, such that $\delta_R(y_n, y) \to 0$. Then the depth function is continuous in $y$ in the sense that,
\begin{eqnarray*}
\lim_{n \to \infty} \te{D}(\nu, y_n) &=& \te{D}(\nu, y).
\end{eqnarray*}
\textbf{(P.6)} \emph{(Uniform continuity in $\nu$)} The depth function is uniformly continuous in terms of the probability measure $\nu$ in the sense that if $(\nu_n)_{n \in \mathbb{N}}$ is a uniformly integrable sequence of probability measures, such that $\nu_n \overset{w}{\to} \nu$. Then, 
\begin{eqnarray*}
\sup_{y \in \mathcal{M}} |\te{D}(\nu_n, y) - \te{D}(\nu, y)| & \to & 0, \quad \quad \te{as } n \to \infty.
\end{eqnarray*}
\subsection{Intrinsic zonoid depth} \label{sec:3.2}
\begin{figure} 
\centering
  \includegraphics[scale=0.37]{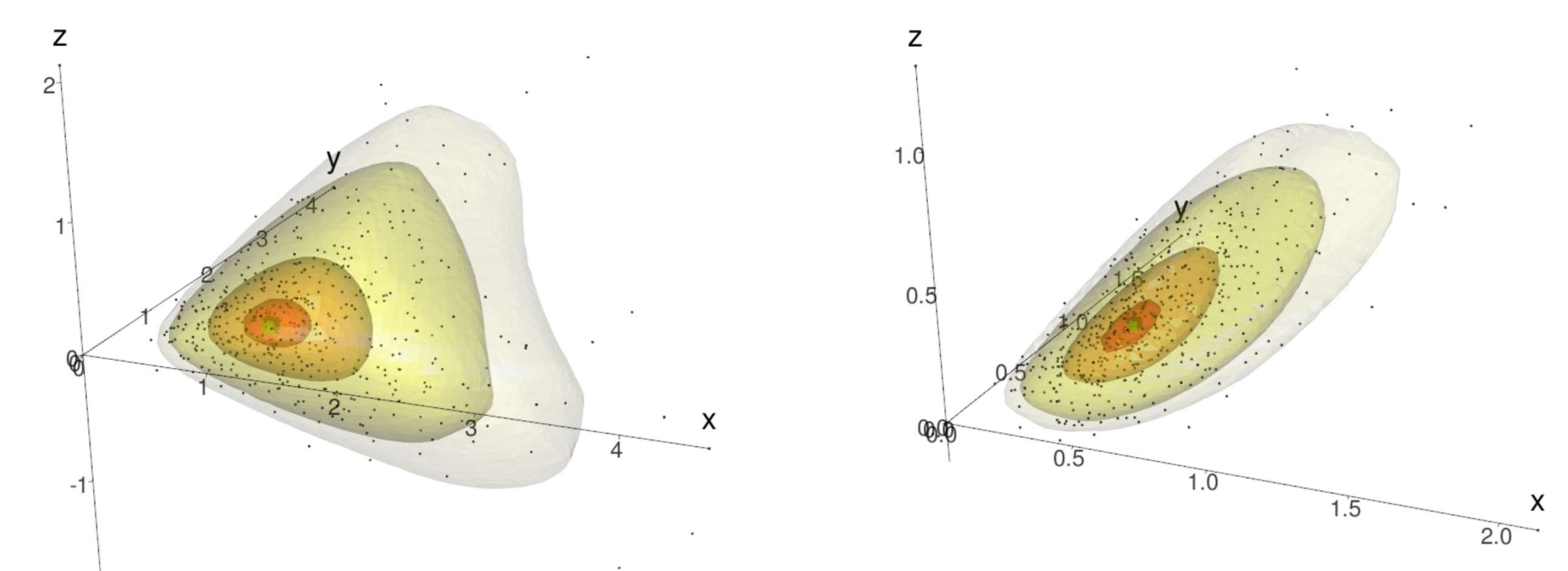}
  \caption{$100(1-\alpha)\%$ intrinsic zonoid depth regions with $\alpha = \{ 0.025, 0.25, 0.75, 0.9 \}$ for random SPD matrices, $(d = 2, n = 500)$, from a Riemannian log-normal distribution (left) and a rescaled Wishart distribution (right), with intrinsic mean $\mu$ shown by the green cube. \label{fig:1}}
\end{figure}

As geodesic convex hulls are well-defined on the Riemannian manifold $(\mathcal{M}, g_R)$, there exist natural manifold generalizations of the simplicial depth or convex hull peeling depth (\cite{L99}) for Euclidean vectors. However, the simplicial depth requires the computation of possibly many convex hulls, which quickly becomes computationally infeasible, especially for higher-dimensional matrices. Instead, we propose a straightforward manifold generalization of another depth measure based on trimmed convex depth regions, the zonoid depth (e.g., \cite{M02}). The intrinsic manifold zonoid depth can be computed with the same tools as the standard zonoid depth for Euclidean vectors and its computation remains efficient, also for higher-dimensional HPD matrices.\\[3mm]
In a Euclidean context, let $\zeta$ be a probability measure on $(\mathbb{R}^d, \mathcal{B}^d)$ with finite first moment, then the zonoid $\alpha$-trimmed region, with $0 < \alpha \leq 1$, is defined as the set,
\begin{eqnarray*}
D_\alpha(\zeta) &:=& \Bigg\{ \int_{\mathbb{R}^d} x w(x)\ d\zeta(x)\, \Big|\, w:\mathbb{R}^d \to \left[0, 1/\alpha\right] \te{measurable, s.t. } \int_{\mathbb{R}^d} w(x)\ d\zeta(x) = 1 \Bigg\}.
\end{eqnarray*}
If $\alpha = 0$, we set $D_0(\zeta) = \mathbb{R}^d$. By \cite[Chapter 3]{M02}, $D_\alpha(\zeta)$ is convex and monotone decreasing in $\alpha$, creating a nested sequence of convex sets for decreasing values $\alpha_1 \geq \ldots \geq \alpha_n$. If $\alpha = 1$, $D_{\alpha}(\zeta)$ consists of the single point $\bs{E}_\zeta[X]$, the Euclidean mean of the distribution $\zeta$.  
The Euclidean zonoid depth of a point $y \in \mathbb{R}^d$ with respect to a distribution $\zeta$ is characterized by the smallest $\alpha$-trimmed region still containing $y$,
\begin{eqnarray*}
\te{ZD}_{\mathbb{R}^d}(\zeta, y) &:=& \sup\left\{ \alpha : y \in D_\alpha(\zeta) \right\}.
\end{eqnarray*}
The zonoid data depth is extended to the Riemannian manifold as follows.
\begin{defn} \emph{(Intrinsic zonoid depth)} \label{def:3.1}
Let $\nu \in P_2(\mathcal{M})$ and let $\zeta_y$ be the probability measure on $(\mathbb{R}^{d^2}, \mathcal{B}(\mathbb{R}^{d^2}))$ of the random variable $\Log_y(X) \in T_y(\mathcal{M}) \cong \mathbb{R}^{d^2}$ as a $d^2$-dimensional random real basis component vector, where $X$ has probability measure $\nu$. The intrinsic zonoid depth of a point $y \in \mathcal{M}$ with respect to the distribution $\nu$ is defined as:
\begin{eqnarray} \label{eq:3.2}
\te{ZD}_{\mathcal{M}}(\nu, y) &:=& \sup\left\{ \alpha : \vec{0} \in D_{\alpha}(\zeta_y) \right\},
\end{eqnarray}
where $\vec{0}$ is a $d^2$-dimensional zero vector, and $D_{\alpha}(\zeta_y)$ is the Euclidean zonoid $\alpha$-trimmed region of the distribution of the normal coordinate vector $\zeta_y$ on $(\mathbb{R}^{d^2}, \mathcal{B}(\mathbb{R}^{d^2}))$. Equivalently, the intrinsic zonoid depth can be written as,
\begin{eqnarray*}
\te{ZD}_{\mathcal{M}}(\nu, y) &=& \sup\left\{ \alpha : y \in D^{\mathcal{M}}_{\alpha}(\nu) \right\},
\end{eqnarray*}
where $D^{\mathcal{M}}_{\alpha}(\nu)$ is the \emph{intrinsic zonoid $\alpha$-trimmed region} defined as, 
\begin{small}
\begin{eqnarray*}
D_\alpha^{\mathcal{M}}(\nu)\ =\ \Bigg\{ y \in \mathcal{M}\, \Big|\, y = \Exp_{y}\left(\int_{\mathcal{M}} \Log_{y}(x) w(x)\: \nu(dx) \right),\ w:\mathcal{M} \to [0, 1/\alpha],  \int_{\mathcal{M}} w(x)\: \nu(dx) = 1 \Bigg\},
\end{eqnarray*}
\end{small}
with $w$ a measurable function.
\end{defn}
\begin{remark} 
Computation of the intrinsic zonoid depth is straightforward via the definition  $\te{ZD}_{\mathcal{M}}(\nu, y) = \te{ZD}_{\mathbb{R}^{d^2}}(\zeta_y, 0)$ and can be calculated directly by the Euclidean zonoid depth as in \cite[Chapter 4]{M02}. Note that if $(e_1,\ldots,e_{d^2})$ is an orthonormal basis of the vector space $(\mathcal{H}, \langle \cdot, \cdot \rangle_F)$, then an orthonormal basis of $(T_y(\mathcal{M}), \langle \cdot, \cdot \rangle_y)$ is simply $(y^{1/2} \ast e_1,\ldots, y^{1/2} \ast e_{d^2})$. In fact, the basis components of $\Log_y(x) \in T_y(\mathcal{M})$ can be computed directly using only an orthonormal basis of $(\mathcal{H}, \langle \cdot, \cdot \rangle_F)$, since $\langle \Log_y(x), y^{1/2} \ast e_i \rangle_y = \langle \Log(y^{-1/2} \ast x), e_i \rangle_F$.
\end{remark}
\begin{theorem} \label{thrm:3.1}
The intrinsic zonoid depth is a proper data depth function in the sense of Section \ref{sec:3.1}, satisfying properties \textbf{P.1}--\textbf{P.4} for distributions in $P_2(\mathcal{M})$. The unique point of maximum depth coincides with the intrinsic mean of the distribution.
\end{theorem}
\noindent In order to show that the continuity properties \textbf{P.5} and \textbf{P.6} also hold for the intrinsic zonoid depth, we need the following lemma. 
\begin{lemma} \label{lem:3.2}
Let $\nu \in P_2(\mathcal{M})$. Then, $\bigcup_{0 < \alpha \leq 1} D_{\alpha}^{\mathcal{M}}(\nu) \: =\: \te{conv}(\nu)$. In particular, for each $y \in \te{conv}(\nu)$, $\te{ZD}_{\mathcal{M}}(\nu, y) > 0$ by definition of the intrinsic zonoid depth.
\end{lemma}
\begin{theorem} \label{thrm:3.3}
The intrinsic zonoid depth is continuous in $y$ as in \textbf{P.5} for $y \in \te{conv}(\nu)$ and $\nu \in P_2(\mathcal{M})$, i.e., if $\delta_R(y_n, y) \to 0$ with $y_n \in \mathcal{M}$ for all $n \in \mathbb{N}$, then,
\begin{eqnarray*}
\lim_{n \to \infty} \te{ZD}_{\mathcal{M}}(\nu, y_n) &=& \te{ZD}_{\mathcal{M}}(\nu, y).
\end{eqnarray*}
The intrinsic zonoid depth is uniformly continuous in $\nu$ as in \textbf{P.6} for $y \in \te{rint}(\te{conv}(\nu))$ and $(\nu_n)_{n \in \mathbb{N}}$ in $P_2(\mathcal{M})$ uniformly integrable. If $\nu_n \overset{w}{\to} \nu$, then,
\begin{eqnarray*}
\sup_{y \in \te{rint}(\te{conv}(\nu))} | \te{ZD}_{\mathcal{M}}(\nu_n, y) - \te{ZD}_{\mathcal{M}}(\nu, y)| & \to & 0, \quad \quad \te{as } n \to \infty.
\end{eqnarray*}
\begin{example} \label{ex:3.1}
In Figure \ref{fig:1}, we display several $100(1-\alpha)\%$ central intrinsic zonoid depth regions for generated i.i.d.\@ samples of $(2 \times 2)$-dimensional SPD matrices $x_1,\ldots,x_{500}$ from a distribution $\nu_\mu \in P_2(\mathcal{M})$ with intrinsic mean $\mu$. Denoting $\nu_{500}$ for the empirical distribution of $x_1,\ldots,x_{500}$, the $100(1-\alpha)\%$ central depth-region $\te{DR}_{1-\alpha}$ is given by the set of SPD matrices:
\begin{small}
\begin{eqnarray*}
\te{DR}_{1-\alpha} &=& \Bigg\{ y \in \te{Re}(\mathbb{P}_{2 \times 2}) \, :\, D(\nu_{500}, y) \geq \beta_*,\ \beta_* := \arg\min_{\beta \in (0,1)} \bigg[ \frac{1}{500} \sum_{i=1}^{500} \bs{1}_{\{ D(\nu_{500}, x_i) \geq \beta\}} \geq 1-\alpha \bigg] \Bigg\},
\end{eqnarray*}
\end{small} 
In the left-hand image, data matrices are sampled from a Riemannian log-normal distribution $\nu_\mu$ as in e.g., \cite{Y12}, with intrinsic mean $\mu$ equal to the identity matrix. That is, $X_i \overset{d}{=} \Exp(\sum_k Z_{ki}e^k)$, with $(Z_{ki})_k \overset{\te{iid}}{\sim} N(0, 1/2)$, where $(e^1,\ldots, e^{4}) \in \mathbb{H}_{2 \times 2}^{4}$ is an orthonormal basis of ($\mathbb{H}_{2 \times 2}, \langle \cdot, \cdot \rangle_F$). In the right-hand image, $\nu_{\mu}$ is a rescaled Wishart distribution with intrinsic mean $\mu = \left(\begin{smallmatrix} 0.5 & 0.25 \\ 0.25 & 0.5 \end{smallmatrix}\right)$, such that $X_i \overset{d}{=} e^{-c(2,8)} W$, with $W \sim W_2^c(8,\mu/8)$ a complex Wishart distribution with $8$ degrees of freedom and $c(d,B) = -\log(B) + \frac{1}{d}\sum_{i=1}^d \psi(B - (d - i))$ the intrinsic bias-correction in \cite[Theorem 5.1]{CvS17}. The $(x,y,z)$-axes in Figure \ref{fig:1} correspond to the three independent components in the symmetric matrix $\left( \begin{smallmatrix} x & z \\ z & y \end{smallmatrix} \right)$.
\end{example}
\end{theorem}
\subsection{Integrated intrinsic zonoid depth} \label{sec:3.3}
A straightforward generalization of the pointwise intrinsic zonoid depth in Definition \ref{def:3.1} to compute the depth of a curve $y(t) \in \mathcal{M}$ with respect to a collection of marginal measures $\nu(t)$ for $t \in \mathcal{I} \subset \mathbb{R}$ is to consider the \emph{integrated} intrinsic zonoid depth given by,
\begin{eqnarray*}
\te{iZD}_{\mathcal{M}}(\nu, y) \ := \  \int_{\mathcal{I}} \te{ZD}_{\mathcal{M}}(\nu(t), y(t))\: dt \ =\ \int_{\mathcal{I}} \sup \left\{ \alpha : \bs{0}_{d \times d} \in D_\alpha(\zeta_{y}(t)) \right\}\: dt,
\end{eqnarray*}
where $\zeta_{y}(t)$ is the probability of the components of the random variable $\Log_{y(t)}(X(t)) \in T_{y(t)}(\mathcal{M}) \cong \mathbb{R}^{d^2}$, such that $X(t)$ has probability measure $\nu(t)$. This is similar to the construction of the modified band depth (MBD) in a functional data context, where the pointwise Euclidean simplicial depths $y(t)$ are integrated over a functional domain $t \in \mathcal{I}$, (\cite{LR09} or \cite{SG12}). The integrated versions of the properties \textbf{P.1} to \textbf{P.6} continue to hold for the integrated intrinsic zonoid depth and are straightforward generalizations of their pointwise analogs.
\begin{theorem} \label{thrm:3.4}
The integrated intrinsic zonoid depth is a proper integrated depth function in the sense of Section \ref{sec:3.1}, satisfying the integrated versions of properties \textbf{P.1}--\textbf{P.4} for collections of marginal distributions $\nu(t) \in P_2(\mathcal{M})$ for $t \in \mathcal{I}$. The unique curve of maximum depth coincides with the curve of pointwise intrinsic means of the marginal distributions.
\end{theorem}
\begin{prop} \label{prop:3.5}
Let $y(t) \in \te{conv}(\nu(t))$, $\nu(t) \in P_2(\mathcal{M})$ and $y_n(t) \in \mathcal{M}$ for each $t \in \mathcal{I}$, such that $y_n(t) \to y(t)$ uniformly in $t$, i.e., $\sup_{t \in \mathcal{I}} \delta_R(y_n(t), y(t)) \to 0$. Then the integrated manifold zonoid depth is continuous in $y$ as in \textbf{P.5} in the sense that,
\begin{eqnarray*}
\lim_{n \to \infty} \te{iZD}_{\mathcal{M}}(\nu, y_n) &=& \te{iZD}_{\mathcal{M}}(\nu, y).
\end{eqnarray*} 
If $y(t) \in \te{rint}(\te{conv}(\nu))$, $(\nu_n(t))_{n \in \mathbb{N}}$ in $P_2(\mathcal{M})$ is a uniformly integrable sequence of measures uniform in $t$, and $\nu_n(t) \overset{w}{\to} \nu(t)$ uniformly in $t$. Then, 
\begin{eqnarray*}
\sup_{y \in \te{rint}(\te{conv}(\nu))} | \te{iZD}_{\mathcal{M}}(\nu_n, y) - \te{iZD}_{\mathcal{M}}(\nu, y)| & \to & 0, \quad \quad \te{as } n \to \infty.
\end{eqnarray*}
Here, $y \in \te{rint}(\te{conv}(\nu))$ means that $y(t) \in \te{rint}(\te{conv}(\nu(t)))$ for each $t \in \mathcal{I}$, and the uniform weak convergence $\nu_n(t) \overset{w}{\to} \nu(t)$ is read as $\sup_{t \in \mathcal{I}} |\mathbb{E}_{\nu_n(t)}[\phi(X)] - \mathbb{E}_{\nu(t)}[\phi(X)]| \to 0$ for every continuous and bounded function $\phi: \mathcal{M} \to \mathbb{R}$.
\end{prop}

\subsection{Geodesic distance depth} \label{sec:3.4}
As a second notion of data depth on the geodesically complete manifold $(\mathcal{M}, g_R)$, we consider the geodesic distance depth, the natural analog on the metric space $(\mathcal{M}, \delta_R)$ of the arc distance depth in \cite{LS92} for data observations on circles and spheres. The geodesic distance depth is straightforward to calculate, also for high-dimensional matrices, as the only required operation is the computation of Riemannian distances between HPD matrices. 
\begin{defn} \emph{(Geodesic distance depth)} \label{def:3.2}
Let $\nu \in P_1(\mathcal{M})$, then the geodesic distance depth of a point $y \in \mathcal{M}$ with respect to the distribution $\nu$ is defined as:
\begin{eqnarray} \label{eq:3.3}
\te{GDD}(\nu, y) &=& \exp\left( - \int_{\mathcal{M}} \delta_R(y, x)\: \nu(dx) \right).
\end{eqnarray}
\end{defn}
\begin{theorem} \label{thrm:3.6}
The geodesic distance depth is a proper data depth function in the sense of Section \ref{sec:3.1}, satisfying \textbf{P.1}--\textbf{P.4} for distributions in $P_1(\mathcal{M})$. The unique point of maximum depth coincides with the intrinsic median of the distribution.
\end{theorem}
\begin{theorem} \label{thrm:3.7}
The geodesic distance depth is continuous in $y$ as in \textbf{P.5} for $y \in \te{cl}(\mathcal{M})$, the closure of $\mathcal{M}$, and $\nu \in P_1(\mathcal{M})$. That is, if $\delta_R(y_n, y) \to 0$ with $y_n \in \mathcal{M}$ for all $n \in \mathbb{N}$, then,
\begin{eqnarray*}
\lim_{n \to \infty} \te{GDD}(\nu, y_n) &=& \te{GDD}(\nu, y).
\end{eqnarray*}
The geodesic distance depth is uniformly continuous in $\nu$ as in \textbf{P.6} for $y \in \mathcal{M}$ and $(\nu_n)_{n \in \mathbb{N}}$ uniformly integrable. If $\nu_n \overset{w}{\to} \nu$, then, 
\begin{eqnarray*}
\sup_{y \in \mathcal{M}} | \te{GDD}(\nu_n, y) - \te{GDD}(\nu, y)| & \to & 0, \quad \quad \te{as } n \to \infty.
\end{eqnarray*}
\end{theorem}
\begin{figure}
\centering
  \includegraphics[scale=0.37]{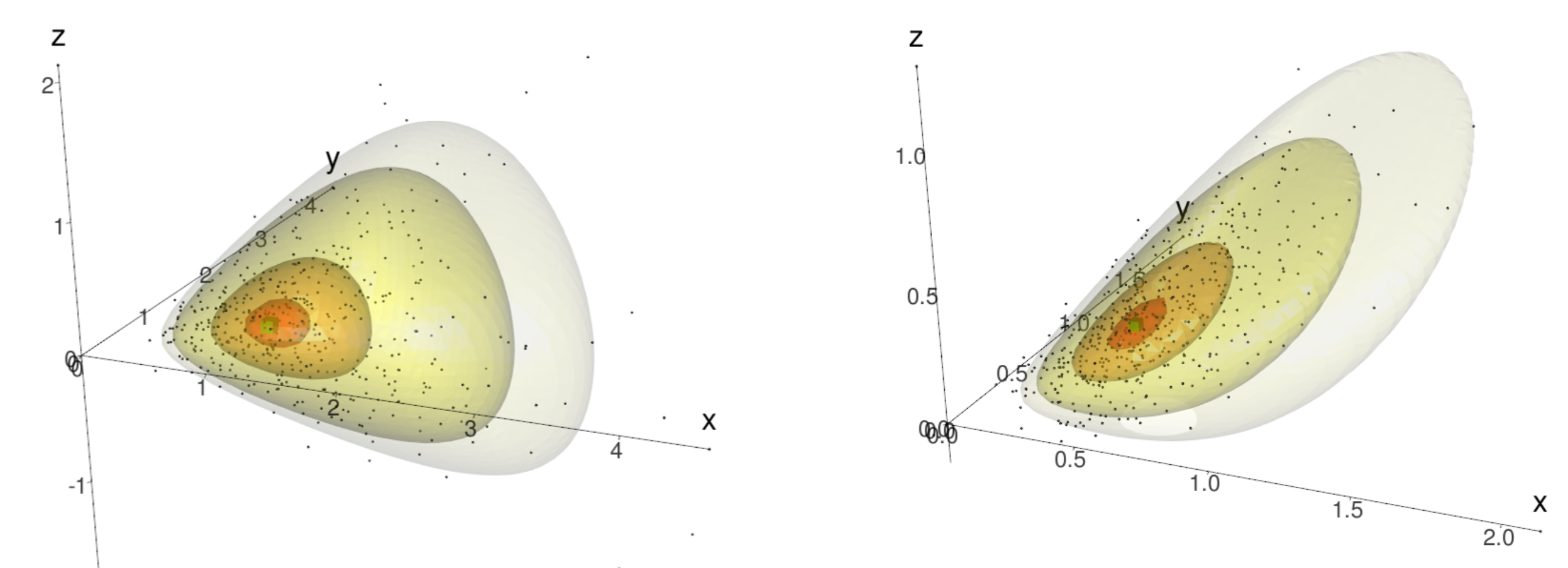}
  \caption{$100(1-\alpha)\%$ central geodesic distance depth regions with $\alpha = \{ 0.025, 0.25, 0.75, 0.9 \}$ for random SPD matrices, $(d = 2, n = 500)$, from a Riemannian log-normal distribution (left) and from a rescaled Wishart distribution (right) as explained in Example \ref{ex:3.1}. \label{fig:2}}
\end{figure}
\begin{remark}
In order to compute the empirical depth $\te{GDD}(\nu_n, y)$ of each observation in a sample $y \in \{x_1,\ldots,x_n\}$ with respect to the empirical distribution $\nu_n$ of the sample $\{x_1,\ldots,x_n\}$, it suffices to compute the $(n \times n)$-dimensional distance matrix with $(i,j)$-th entry $\delta_R(x_i, x_j)$. This matrix is fully determined by $n(n-1)/2$ components, as the diagonal entries are zero and $\delta_R(x_i, x_j) = \delta_R(x_j, x_i)$. In particular, in online applications where the depths need to be updated each time a new observation enters the database, we simply add one extra column and row to the distance matrix and update the depth values.  
\end{remark}
\begin{remark}
A third notion of data depth on the Riemannian manifold $(\mathcal{M}, g_R)$, closely related to the geodesic distance depth, is the \emph{intrinsic spatial depth}. This is the natural manifold generalization of the spatial depth in \cite{VZ00} or \cite{S02}. For a distribution $\nu \in P_1(\mathcal{M})$ and a point $y \in \mathcal{M}$, the intrinsic spatial depth is given by:
\begin{eqnarray*}
\te{SD}(\nu, y) \ =\ 1 - \Bigg\Vert \int_{\mathcal{M}} \frac{\Log_y(x)}{\delta_R(y,x)}\ \nu(dx) \Bigg\Vert_{y} \ = \ 1 - \Bigg\Vert \int_{\mathcal{M}} \frac{\Log(y^{-1/2} \ast x)}{\delta_R(y,x)}\ \nu(dx) \Bigg\Vert_F.
\end{eqnarray*}
The intrinsic spatial depth attains its maximum value $\te{SD}(\nu, m) = 1$ at the intrinsic median, since $\bs{E}_{\nu}\left[ \frac{\Log_m(x)}{\delta_R(m,x)} \right] = \bs{0}$ by definition of the intrinsic median, and the depth is lower bounded by zero, which is a direct consequence of the triangle inequality combined with the fact that $\Vert \Log_y(x) \Vert_y = \delta_R(y, x)$. The intrinsic spatial depth is closely associated to the geodesic distance depth in the sense that it is based on the gradient of the distance function, i.e., the gradient of $f_x(y) = \delta_R(y,x)$ for fixed $x$ is given by $\te{grad} f_x(y) = \frac{\Log_y(x)}{\delta_R(y,x)}$, see \cite{F09}. 
\end{remark}
\begin{figure}
\centering
  \includegraphics[scale=0.37]{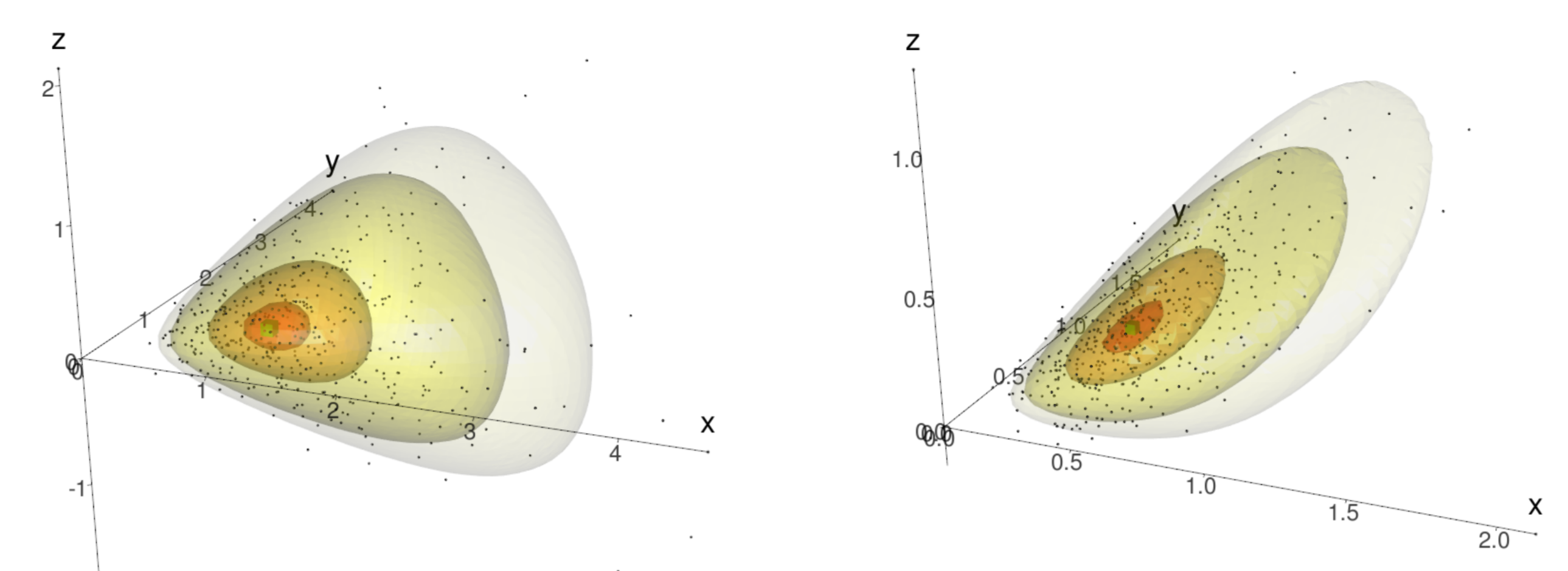}
  \caption{$100(1-\alpha)\%$ central intrinsic spatial depth regions with $\alpha = \{ 0.025, 0.25, 0.75, 0.9 \}$ for random SPD matrices, $(d = 2, n = 500)$, from a Riemannian log-normal distribution (left) and from a rescaled Wishart distribution (right) as outlined in Example \ref{ex:3.1}. \label{fig:3}}
\end{figure}
\subsection{Integrated geodesic distance depth}
In order to generalize the pointwise geodesic distance depth to the depth of a curve $y(t) \in \mathcal{M}$, with respect to a collection of marginal measures $\nu_t = \nu(t)$ for $t \in \mathcal{I} \subset \mathbb{R}$, we replace the pointwise expected distance in Definition \ref{def:3.2} by an integrated expected distance as:
\begin{eqnarray*}
\te{iGDD}(\nu, y) &=& \exp\left( - \int_{\mathcal{I}}\int_{\mathcal{M}} \delta_R(y(t), x)\: \nu_t(dx)\ dt \right).
\end{eqnarray*}
The integrated versions of the properties \textbf{P.1} to \textbf{P.6} continue to hold for the integrated geodesic distance depth and are straightforward generalizations of their pointwise analogs as in the case of the integrated intrinsic zonoid depth.
\begin{theorem} \label{thrm:3.8} 
The integrated geodesic distance depth is a proper function depth function in the sense of Section \ref{sec:3.1}, satisfying the integrated versions of properties \textbf{P.1}--\textbf{P.4} for collections of marginal distribution $\nu(t) \in P_1(\mathcal{M})$ for $t \in \mathcal{I}$. The unique curve of maximum depth coincides with the curve of pointwise intrinsic medians of the marginal distributions.
\end{theorem}
\begin{prop} \label{prop:3.9}
Let $y(t) \in \te{cl}(\mathcal{M})$ and $\nu(t) \in P_1(\mathcal{M})$ for each $t \in \mathcal{I}$, such that $y_n(t) \to y(t)$ uniformly in $t$, i.e., $\sup_{t \in \mathcal{I}} \delta_R(y_n(t), y(t)) \to 0$. Then the integrated geodesic distance depth is continuous in $y$ as in \textbf{P.5} in the sense that, 
\begin{eqnarray*}
\lim_{n \to \infty} \te{iGDD}(\nu, y_n) &=& \te{iGDD}(\nu, y).
\end{eqnarray*}
If $y(t) \in \mathcal{M}$, $(\nu_n(t))_{n \in \mathbb{N}}$ in $P_1(\mathcal{M})$ is a uniformly integrable sequence of measures uniform in $t$, and $\nu_n(t) \overset{w}{\to} \nu(t)$ uniformly in $t$. Then,
\begin{eqnarray*}
\sup_{y \in \mathcal{M}} | \te{iGDD}(\nu_n, y) - \te{iGDD}(\nu, y)| & \to & 0, \quad \quad \te{as } n \to \infty,
\end{eqnarray*}
where $y \in \mathcal{M}$ is read as $y(t) \in \mathcal{M}$ for each $t \in \mathcal{I}$.
\end{prop}
\section{Aspects of robustness and efficiency} \label{sec:4}

\begin{figure}
\centering
\begin{subfigure}{0.49\linewidth}
  \centering
  \includegraphics[scale=0.37]{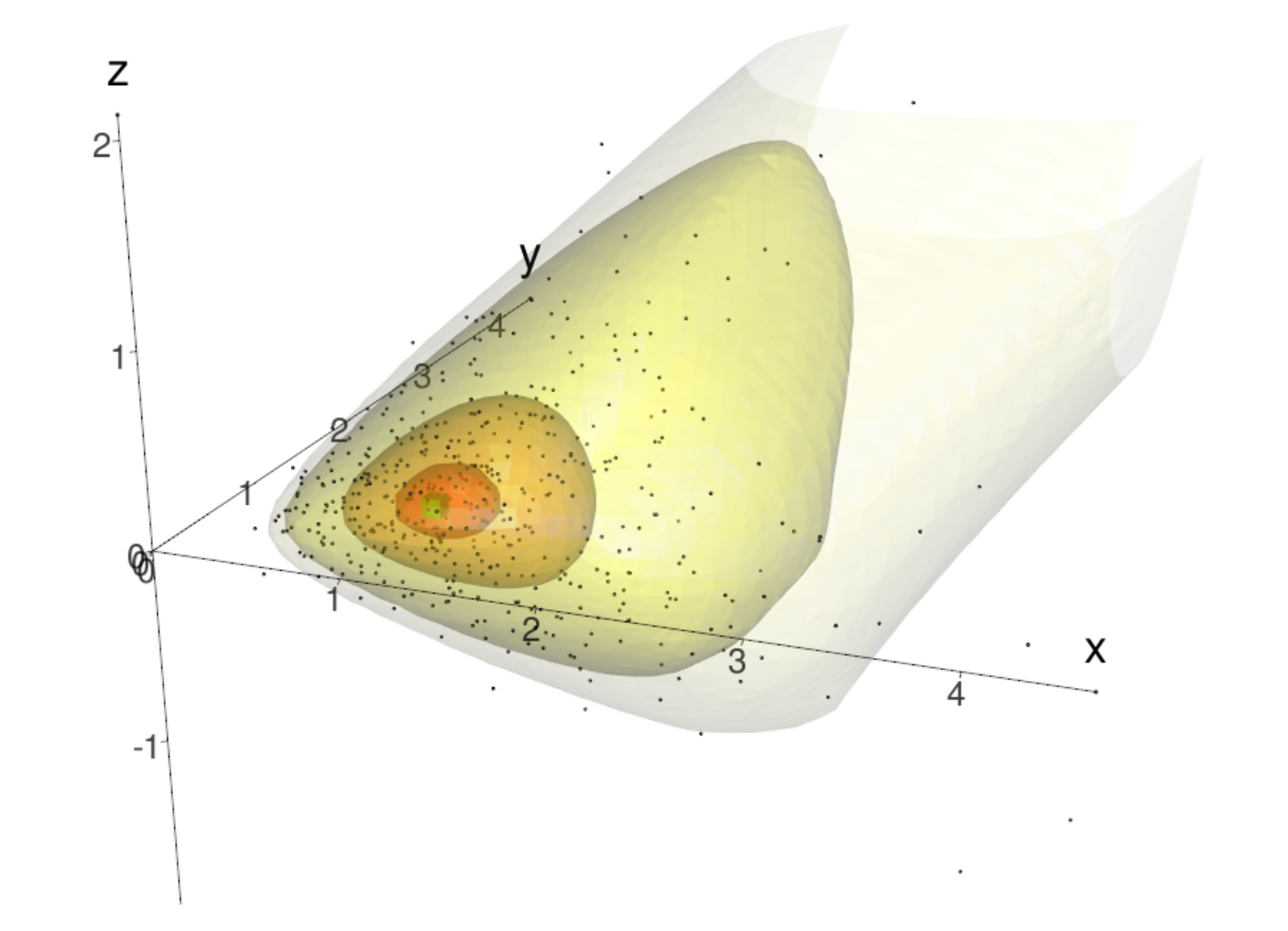}
\end{subfigure}
\begin{subfigure}{0.49\linewidth}
  \centering
  \includegraphics[scale=0.38]{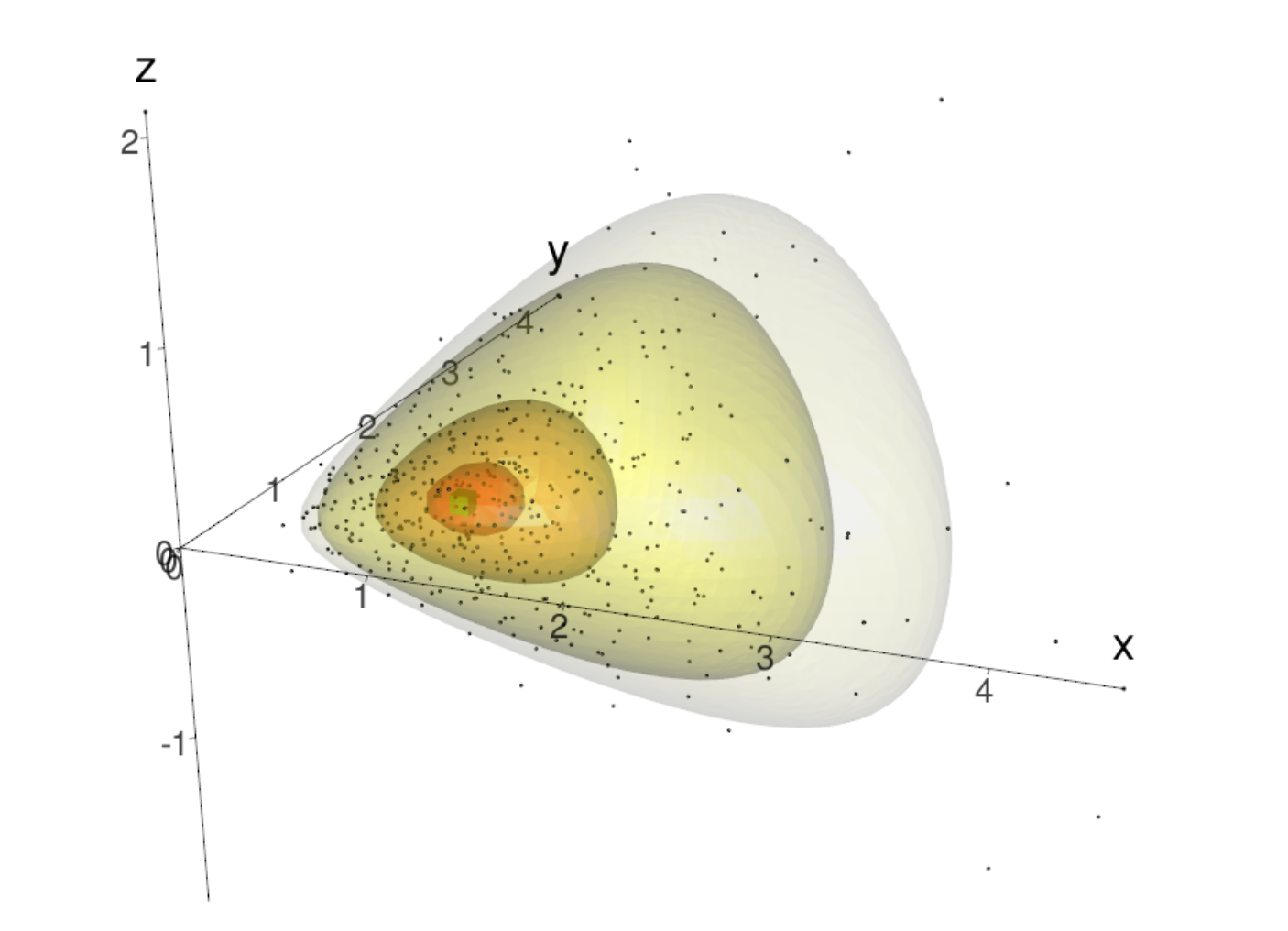}
\end{subfigure}
\caption{$100(1-\alpha)\%$ intrinsic zonoid (left) and geodesic distance (right) central depth regions with $\alpha = \{ 0.025, 0.25, 0.75, 0.9 \}$ for random SPD matrices, $(d = 2, n = 501)$, from a Riemannian log-normal distribution as detailed in Example \ref{ex:3.1} contaminated by a single large SPD matrix with components $(x,y,z) = (10^4, 10^4, 9999)$. \label{fig:4}}
\end{figure}

\paragraph{Depth-median breakdown.} An intuitive measure of robustness of the intrinsic depth functions is given by their breakdown points according to \cite{HRRS86}. In order to assess the robustness of the depth functions, a first step is to compute the breakdown point of the location estimator that maximizes the depth, i.e., the \emph{depth-median}, as in \cite{DG92} or \cite{LS92}, which we explain as follows. Let $X^{(n)} = \{x_1,\ldots,x_n\} \in \mathcal{M}^n$ be an initial set of HPD observations and let $Y^{(m)} = \{ y_1,\ldots,y_m\} \in \mathcal{M}^m$ be a set of contaminating HPD observations. Denote $Z^{(n,m)} = X^{(n)} \cup Y^{(m)}$ and consider the --not necessarily in-sample-- depth-median $T_D(Z^{(n,m)}) = \min_{y \in \mathcal{M}} D(y, \nu_{n,m})$, with $\nu_{n,m}$ the empirical distribution of $Z^{(n,m)}$. The breakdown point of the depth-median is the smallest fraction of arbitrarily large contaminating observations that breaks down the estimator:
\begin{eqnarray} \label{eq:4.1}
\epsilon_1(X) &=& \left\{ \frac{m}{m + n} \ :\ \sup_{Y^{(m)}} \Vert \Log(T_D(Z^{(n,m)})) \Vert_F = \infty \right\}.
\end{eqnarray}
Note that $\Vert \Log(x) \Vert_F = \delta_R(x, \te{Id})$, such that $\Vert \Log(x) \Vert_F < \infty$ for all $x \in \mathcal{M}$, and $\Vert \Log(x) \Vert_F = \infty$ if $x$ is a singular matrix lying on the boundary of the metric space $(\mathcal{M}, \delta_R)$. The breakdown point of the depth-median for the intrinsic zonoid depth is $\epsilon_1(X) = 1/(n+1)$ as the depth-median coincides with the sample intrinsic mean and it requires only a single large contaminating observation to make the sample intrinsic mean arbitrarily large. The intrinsic zonoid depth-median is therefore not robust against outlying observations in terms of the depth-median breakdown point, analogous to the Euclidean case, as discussed in \cite{M02}. For the geodesic distance depth, the depth-median coincides with the intrinsic median and the intrinsic median in a geodesically complete manifold is known to have maximum breakdown point $\epsilon_1(X) = 1/2$, as shown in \cite[Theorem 2]{F11}. 
\paragraph{Simultaneous depth-rank breakdown.} The above definition of the breakdown point gives us an intuitive measure of robustness for the depth-median. However, it does not tell us how robust the depth function is with respect to the depth-ranked observations in the sample itself. As a more informative robustness measure, we study the breakdown point simultaneously over all the depth-ranked observations in an initial sample of size $n$. Let us write $z^{(n,m)}_{[i]}$ for the $i$-th center-to-outward order statistic (or $i$-th depth-ranked observation) with respect to a given depth measure. The simultaneous breakdown point is the smallest fraction of arbitrarily large contaminating observations that breaks down \emph{at least one} of the first $n$ depth-ranked observations:
\begin{eqnarray} \label{eq:4.2}
\epsilon_2(X) &=& \left\{ \frac{m}{n + m} \ :\ \max_{i=1,\ldots,n} \sup_{Y^{(m)}} \Vert \Log(z_{[i]}^{(n,m)}) \Vert_F = \infty \right\}.
\end{eqnarray}
For the intrinsic zonoid depth, if we break ties by assigning the same rank to each observation with equal depth, the simultaneous breakdown point is $\epsilon_2(X) = 1/(n+1)$. If we break ties by assigning increasing ranks based on increasing Frobenius norms $\Vert \Log(z_i^{(n,m)}) \Vert_F$, then the simultaneous breakdown point is $\epsilon_2(X) = 2/(n+2)$. This is illustrated as follows. Let $y_1$ be a first contaminating observation with $\Vert \Log(y_1)\Vert_F > N_M$, such that $\Vert \Log(\widebar{Z}^{(n,1)})\Vert_F > M$, where $\widebar{Z}^{(n,1)}$ denotes the intrinsic mean of the contaminated sample $Z^{(n,1)}$. Assuming without loss of generality that $\Vert \Log(x_i) \Vert_F \ll N_M$ for each $i=1,\ldots,n$, the contaminating observation $y_1$ will be assigned depth-rank $n+1$ and the first $n$ depth-ranked observations do not break down. Let $y_2 = \widebar{Z}^{(n,1)}$ be a second contaminating observation, then $y_2$ has maximum depth by Theorem \ref{thrm:3.1}, and thus $z^{(n,2)}_{[1]} = y_2$. Since we can choose $N_M > 0$, such that $\Vert \Log(y_2) \Vert_F > M$ for any $M > 0$, it follows that $\epsilon_2(X) = 2/(n+2)$. 
\begin{prop} \label{prop:4.1}
For the geodesic distance depth, the depth-ranked observations have maximum simultaneous breakdown point $\epsilon_2(X) = 1/2$ equal to the median breakdown point $\epsilon_1(X)$.
\end{prop}
\noindent The above proposition asserts that if we observe a number of (large) contaminating observations $m$ smaller than the initial sample size $n$, the geodesic distance depth will assign the contaminating observations to the ranks $n+1,\ldots,n+m$. The depth-rankings with respect to the geodesic distance depth are therefore highly robust against arbitrarily large contaminating observations, in contrast to the intrinsic zonoid depth-rankings, also illustrated in Figure \ref{fig:4}. 
\begin{figure}[ht]
\centering
\begin{subfigure}{0.327\linewidth}
  \centering
  \includegraphics[scale=0.55]{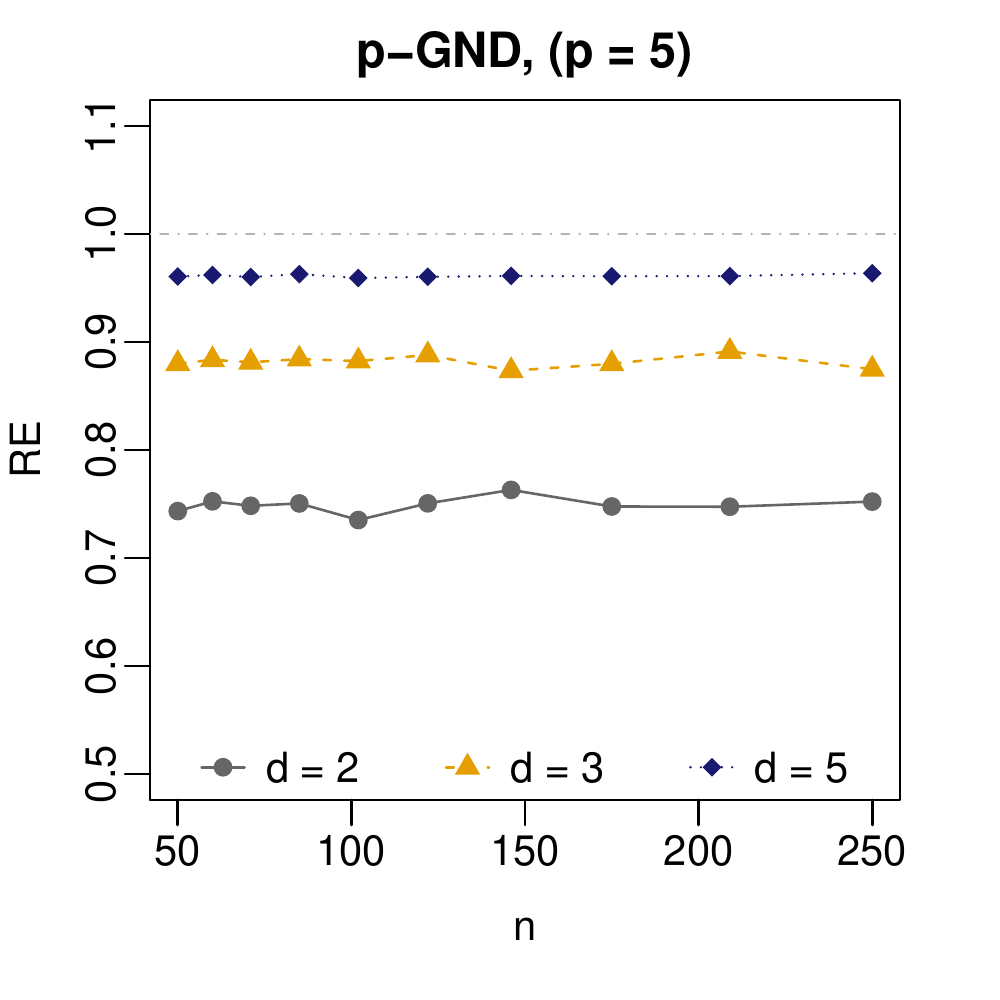}
\end{subfigure}
\begin{subfigure}{0.327\linewidth}
  \centering
  \includegraphics[scale=0.55]{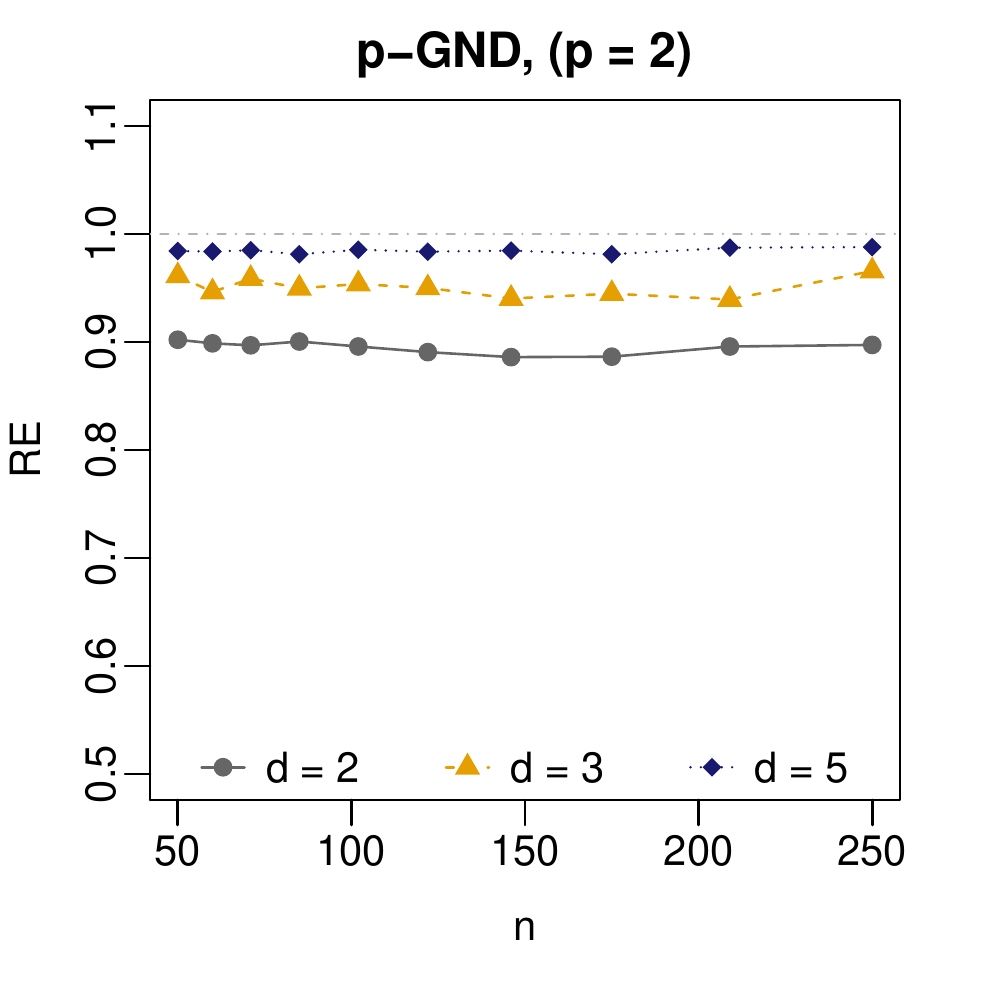}
\end{subfigure}
\begin{subfigure}{0.327\linewidth}
  \centering
  \includegraphics[scale=0.6]{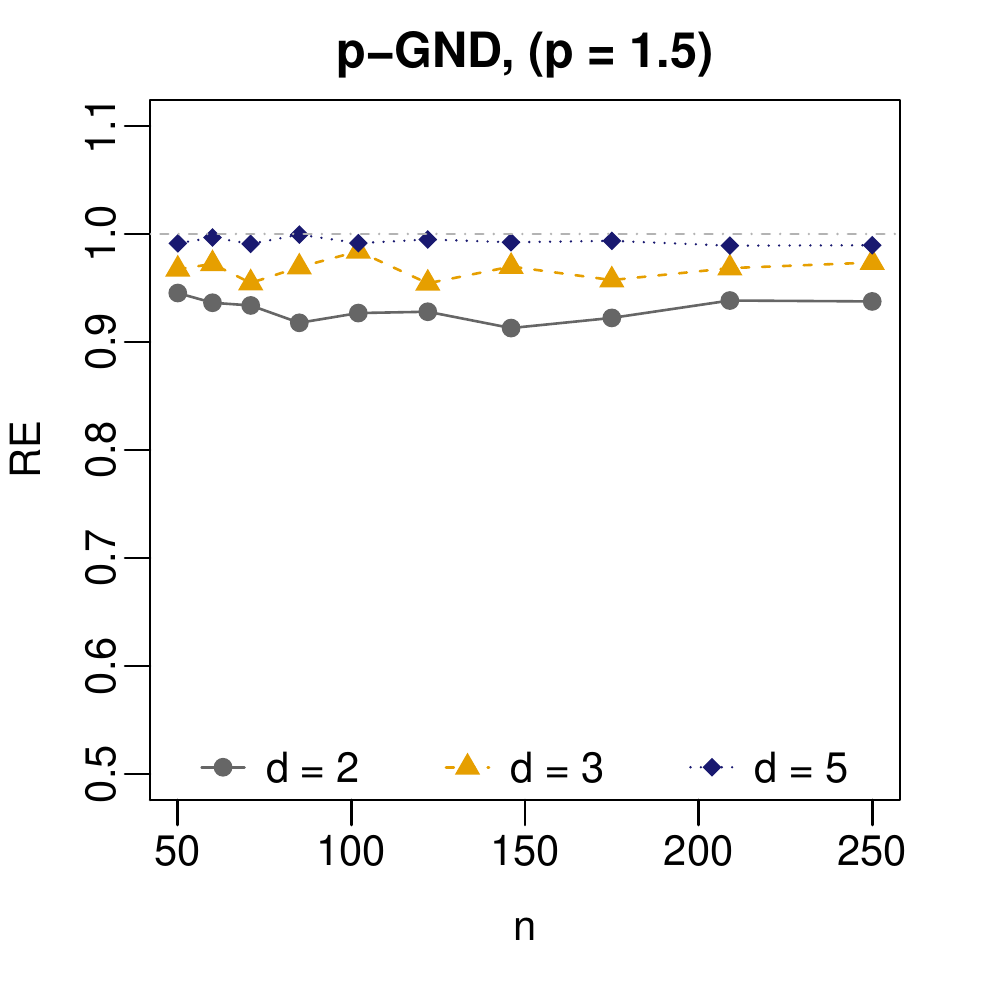}
\end{subfigure}
\caption{Average Riemannian efficiency of the geodesic distance depth-median relative to the intrinsic zonoid depth-median based on $5000$ i.i.d.\@ samples of $(d \times d)$-dimensional HPD matrices of size $n$, with data generation ranging from light- to heavy-tailed $p$-GNDs.}
\label{fig:5}
\end{figure}
\begin{example}
The above depth measures share the same robustness properties in terms of their depth-median and simultaneous depth-rank breakdown point. In general, this does not have to be the case. For instance, consider the simplicial or convex hull peeling depth on the real line (e.g.,\@ \cite{L99}), which are highly robust in terms of their depth-median breakdown point $\epsilon_1(X) = 1/2$, as argued in \cite{C95} for the simplicial depth. In contrast, both data depths have simultaneous breakdown points $\epsilon_2(X) \leq 2/(n+2)$ as two well-placed large contaminating observations $y_1,y_2 \in \mathbb{R}$ can ensure that $\Vert z_{[n]}^{(n,m)} \Vert > M$ for any $M > 0$. 
\end{example}
\begin{remark}
The definitions of the depth-median and simultaneous breakdown points for the integrated depth functions are straightforward generalizations of the pointwise definitions above and it is easily verified that the breakdown points for the integrated depth functions coincide with their pointwise analogs. 
\end{remark}
\paragraph{Depth-median efficiency.}
The robustness of the depth functions may result in a loss of efficiency of the depth-median as an intrinsic location estimator on the Riemannian manifold $(\mathcal{M}, g_R)$. Figure \ref{fig:5} displays the relative efficiency of the geodesic distance depth-median $\hat{\mu}_{\te{GDD}}$, (i.e., the intrinsic median), relative to the intrinsic zonoid depth-median $\hat{\mu}_{\te{ZD}}$, (i.e., the intrinsic mean), in terms of the Riemannian mean squared error. That is,
\begin{eqnarray*}
\te{RE}(\hat{\mu}_{\te{ZD}}, \hat{\mu}_{\te{GDD}}) &=& \frac{\bs{E}_{\nu}[\delta_R(\hat{\mu}_{\te{GDD}}(\bs{X}), \mu)^2]}{\bs{E}_{\nu}[\delta_R(\hat{\mu}_{\te{ZD}}(\bs{X}), \mu)^2]},
\end{eqnarray*}
The depth-medians are computed from simulated samples $\bs{X} = X_1,\ldots,X_n \overset{\te{iid}}{\sim} \nu^p_{\te{Id}}$, where $\nu^p_{\te{Id}} \in P_2(\mathcal{M})$ is a centrally symmetric distribution, such that the intrinsic mean and median coincide and are equal to the identity matrix. In particular, $X_i \overset{d}{=} \Exp( \sum_k Z_k e_k )$, where $(e^1,\ldots, e^{d^2}) \in \mathbb{H}_{d \times d}^{d^2}$ is an orthonormal basis of ($\mathbb{H}_{d \times d}, \langle \cdot, \cdot \rangle_F$), and $(Z_k)_k$ are i.i.d.\@ random variables from a $p$-generalized normal distribution (\cite{SGB09}), with mean zero and standard deviation $\sigma_p = p^{1/p} \sqrt{\Gamma(3/p)/\Gamma(1/p)}$, such that $\sigma_2 = 1$. The family of $p$-generalized normal distributions ($p$-GNDs) allows us to generate tail behavior that is either heavier $(p < 2)$ or lighter $(p > 2)$ than the normal distribution. For $p=2$, the $p$-GND coincides with the normal distribution. As shown in Figure \ref{fig:5}, for random variables generated from a light-tailed $p$-GND ($p=5$ and $p=2$ and in particular small dimensions $d$), the intrinsic zonoid depth regions are better centered around the population mean of the generating distributions than the geodesic distance depth regions; for a heavier-tailed $p$-GND ($p = 1.5$), the efficiency gain of the intrinsic zonoid depth-median relative to the geodesic distance depth-median diminishes.
\subsection{Computational effort}
\begin{figure}
\centering 
\begin{subfigure}{0.49\linewidth}
  \centering
  \includegraphics[scale=0.58]{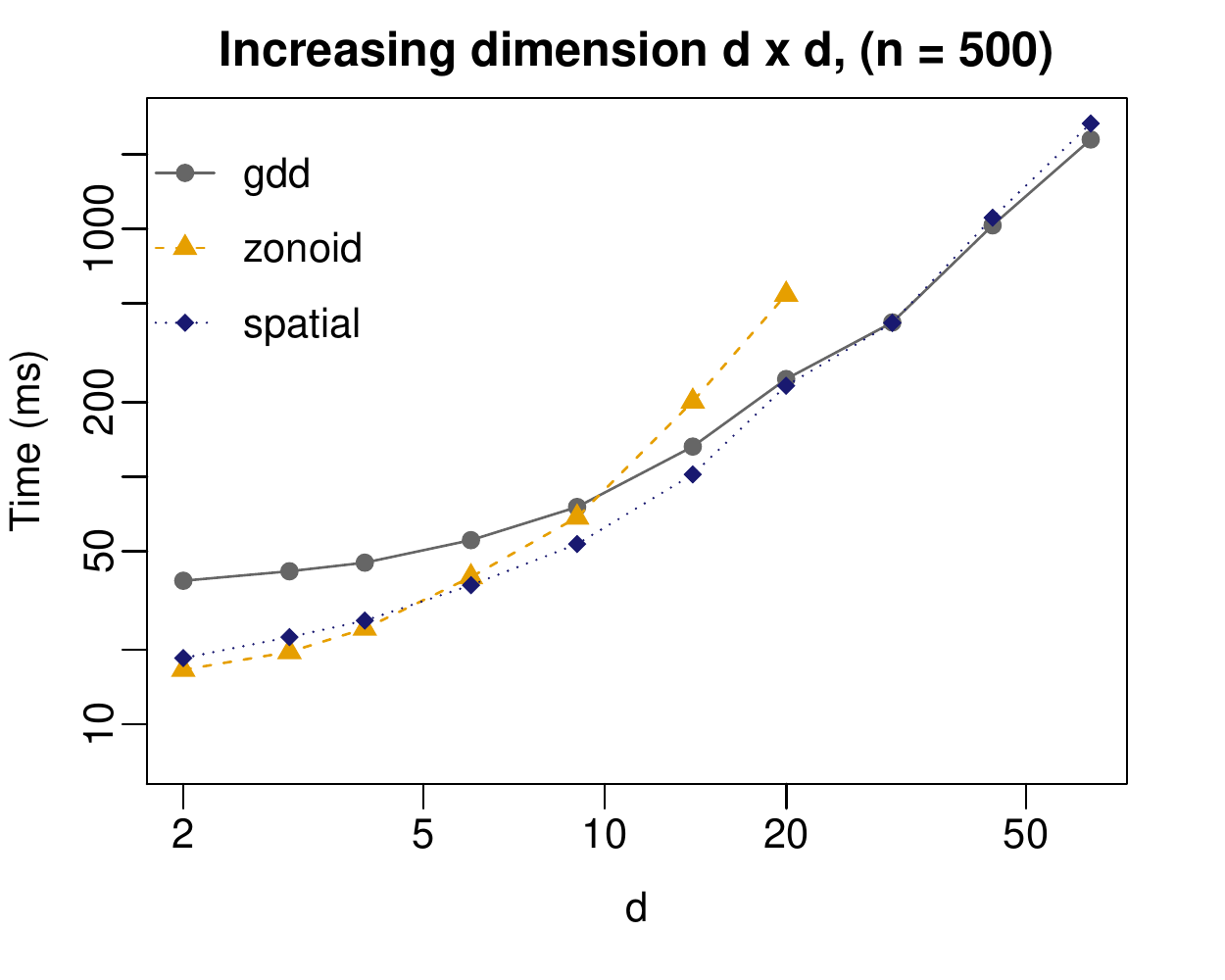}
\end{subfigure}
\begin{subfigure}{0.49\linewidth}
  \centering
  \includegraphics[scale=0.58]{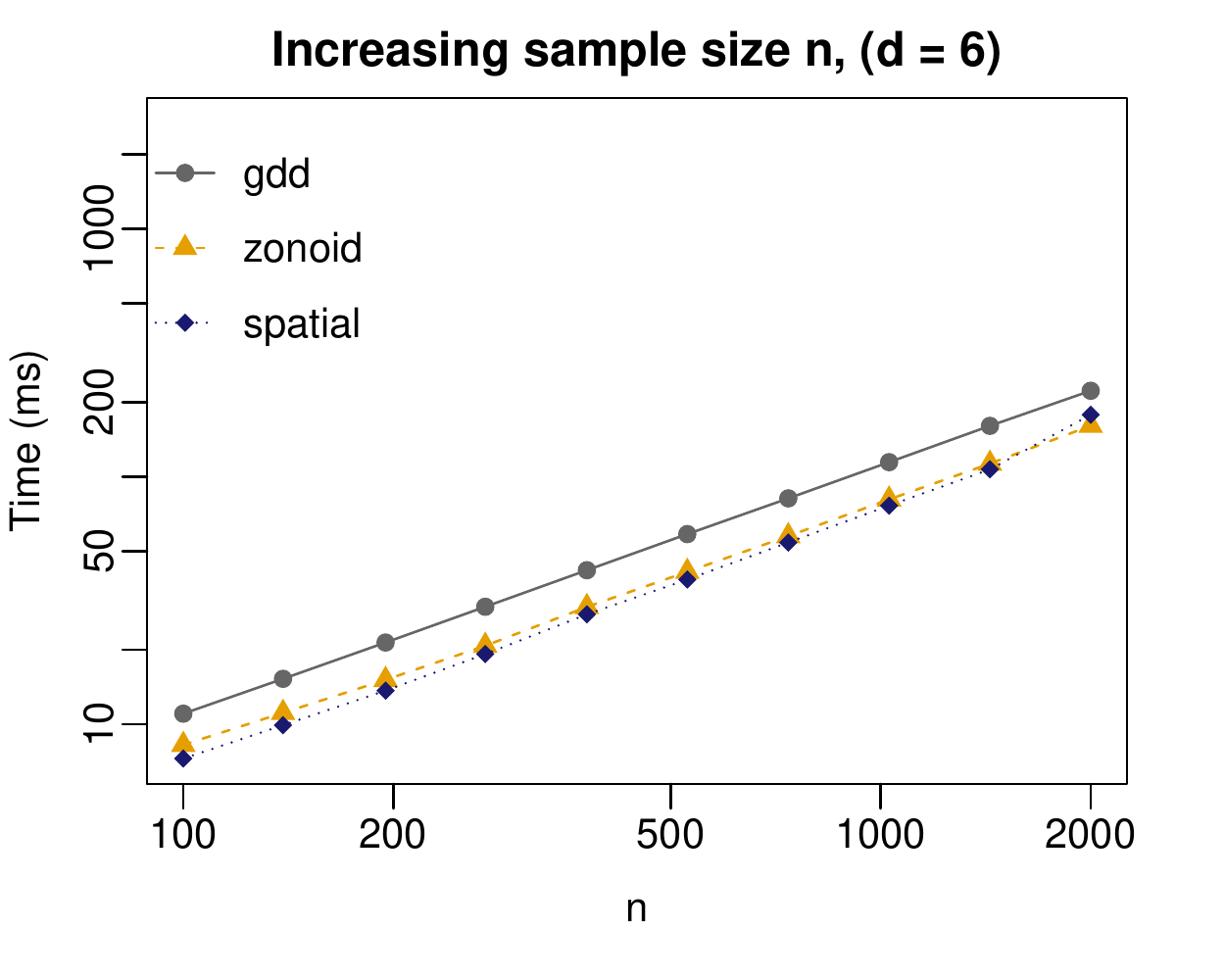}
\end{subfigure}
\caption{Intrinsic zonoid, geodesic distance and intrinsic spatial depth median computation times in milliseconds (ms). \label{fig:6}}
\end{figure}

To demonstrate the computational effort of the depth calculations in practice, Figure \ref{fig:6} displays median computation times in milliseconds (single-core Intel Xeon E5-2650, 2.40Ghz) of the intrinsic depths of a single $(d \times d)$-dimensional HPD matrix with respect to a sample of $n$ HPD matrices calculated with the function \texttt{pdDepth()} in the accompanying R-package \texttt{pdSpecEst}, (including the intrinsic spatial depth computation times). On the left, the sample size is fixed at $n = 500$, and on the right the matrix-dimensions are fixed at $d = 6$. The displayed times are the median computation times of $100$ depth calculations for $100$ simulated samples, i.e., a total of $10^4$ depth calculations per scenario. The intrinsic zonoid depth requires that $d^2 < n$ and for this reason there are several missing values in the left-hand image. Changing the default affine-invariant metric in the intrinsic depth computations to e.g., the Log-Euclidean, Cholesky, root-Euclidean or Euclidean metric --all are available in the function \texttt{pdDepth()}-- the depth computation times are either similar or faster than the times displayed in Figure \ref{fig:6}.
\section{Application: Confidence sets for HPD matrices} \label{sec:5}
As an illustrating application of the intrinsic depth functions, we construct intrinsic matrix confidence regions in the space of HPD or SPD matrices, such as confidence regions for estimated covariance or spectral density matrices. In the context of spectral density matrix estimation, a common approach is to construct asymptotic or bootstrapped confidence regions individually for each element of the spectral matrix, as demonstrated in \cite{DG04} or \cite{FvS14} among others. Although this is a suitable approach to assess the variability of the estimator in each of the individual matrix components, this does not allow for the construction of simultaneous confidence regions across matrix elements, as the combined elementwise confidence intervals do not take the positive definite constraints of the full matrix object into account. In contrast, the intrinsic depth regions provide a natural way to construct joint matrix confidence regions taking into account the geometric constraints of the target space. This is illustrated by the construction of depth-based confidence regions for the intrinsic mean of a sample of i.i.d.\@ HPD random matrices.\\[3mm]
Consider $X_1,\ldots,X_n \overset{\te{iid}}{\sim} \nu_\mu$, with $\nu_{\mu} \in P_2(\mathcal{M})$ centered around a population intrinsic mean $\mu \in \mathcal{M}$. Denote $\bar{m}$ for the sample intrinsic mean, i.e., $\bar{m} := \arg\min_y \sum_{i=1}^n \delta_R(y, X_i)^2$, then the intrinsic central limit theorem in \cite[Proposition 11]{S15} tells us that,
\begin{eqnarray*}
\sqrt{n}\, \Log_{\mu}(\bar{m}) & \overset{d}{\to} & Z,\quad \te{as } n \to \infty,
\end{eqnarray*}
where $Z$ is a random Hermitian matrix, such that $Z \overset{d}{=} \sum_i z_i e^i$, with $(z_1,\ldots,z_{d^2})' \sim N_{d^2}(\bs{0}, \Lambda)$ and $(e^1,\ldots,e^{d^2})$ an orthonormal basis of $T_{\mu}(\mathcal{M})$ equipped with the associated metric $\langle \cdot, \cdot \rangle_\mu$.\\[3mm]
To cast this into a standard Euclidean framework, the Euclidean logarithmic map is given by $\Log_\mu(\bar{m}) = \bar{m} - \mu$. If $\sqrt{n}(\bar{m} - \mu) = Z$ for some fixed matrices $\bar{m}, \mu, Z$, then $\mu = \bar{m} - \frac{1}{\sqrt{n}} Z$, and in the random setting, the construction of asymptotic confidence sets for $\mu$ is straightforward based on an estimate $\bar{m}$ and knowledge of the limiting distribution of $Z$. In a curved Riemannian manifold, if $\sqrt{n} \Log_\mu(\bar{m}) = Z$, with $\bar{m},\mu,Z$ fixed, then in general $\mu \neq \Exp_{\bar{m}}\big(-\frac{1}{\sqrt{n}} Z\big)$. Instead, $\mu = \Exp_{\bar{m}}\big(-\frac{1}{\sqrt{n}} \widetilde{Z}_\mu\big)$, where $\widetilde{Z}_\mu$ is the parallel transport of the matrix $Z$ from the tangent space $T_\mu(\mathcal{M})$ at $\mu$ to the tangent space $T_{\bar{m}}(Z)$ at $\bar{m}$. In the Euclidean setting $\widetilde{Z}_\mu = Z$, as the parallel transport in a Euclidean or flat space equals the identity map, but on the Riemannian manifold $(\mathcal{M}, g_R)$ the parallel transport is nontrivial due to the nonzero curvature of the space and it depends on the unknown population mean $\mu$. One working solution is to approximate the parallel transport using a plug-in estimator for $\mu$, such as $\bar{m}$, in which case the parallel transport is approximated by the identity map. Another approach that is considered here, is to construct approximate confidence sets for the intrinsic mean through resampling, which does not require knowledge of the population mean $\mu$. That is, (i) generate bootstrap intrinsic sample means $\bar{m}^*_1,\ldots, \bar{m}^*_B$ by resampling with replacement from $X_1,\ldots,X_n$, (ii) define a percentile $100(1-\alpha)\%$ confidence region for $\mu$ in the same fashion as \cite{YS97} or \cite{WL12} through the trimmed depth-region:
\begin{small}
\begin{eqnarray}
\te{CR}_{1-\alpha}(\bs{X}) &=& \left\{ \theta \in \mathcal{M}\, :\, D(\theta, \bar{\nu}^*_B) \geq \beta_*,\ \beta_* := \arg\min_{\beta \in (0,1)} \bigg[ \frac{1}{B}\sum_{b=1}^B \bs{1}\{ D(\bar{m}^*_b, \bar{\nu}^*_B) \geq \beta \} \geq 1-\alpha \bigg]  \right\},
\nn
\label{eq:5.1} 
\end{eqnarray}
\end{small}
where $\bar{\nu}^*_B$ is the empirical distribution of $\bar{m}_1^*,\ldots,\bar{m}^*_B$. First-order convergence of the percentile confidence regions to the asymptotically correct confidence regions, as $n$ and $B$ tend to infinity, follows in the same fashion as in \cite{YS97}. The proof relies on the uniform continuity property \textbf{P.6}, satisfied by both the intrinsic zonoid and geodesic distance depth. 
\begin{table}[ht]
\centering
\begin{footnotesize}
\begin{tabular}{l ccc c ccc}
\toprule
&&&&&&& \\[-4mm]
& \multicolumn{3}{c}{\textbf{Intrinsic zonoid depth}, $(n = 100, d = 2)$} & \ & \multicolumn{3}{c}{\textbf{Geodesic distance depth}, $(n = 100, d = 2)$} \\
\cmidrule(lr{0.9em}){2-4} \cmidrule(lr{0.9em}){6-8} \\[-4mm]
\textbf{5-GND} & Ave.-$\beta_*$ & Ave.-Size (SE$\times 10^{-5}$) & Coverage & \ & Ave.-$\beta_*$ & Ave.-Size (SE$\times 10^{-5}$) & Coverage \\[2mm]
$80\%$-CR & 0.0181 & 0.171 (0.78) & \textbf{0.760} & \ & 0.810 & 0.171 (0.21) & \textbf{0.805} \\
$90\%$-CR & 0.0064 & 0.196 (0.95) & \textbf{0.889} & \ & 0.794 & 0.195 (0.25) & \textbf{0.901} \\
$95\%$-CR & 0.0023 & 0.214 (1.20) & \textbf{0.935} & \ & 0.780 & 0.216 (0.30) & \textbf{0.957} \\[1mm]
\midrule\\[-4mm]
\textbf{2-GND} & Ave.-$\beta_*$ & Ave.-Size (SE$\times 10^{-5}$) & Coverage & \ & Ave.-$\beta_*$ & Ave.-Size (SE$\times 10^{-5}$) & Coverage \\[2mm]
$80\%$-CR & 0.0181 & 0.205 (1.71) & \textbf{0.796} & \ & 0.775 & 0.207 (0.61) & \textbf{0.825} \\
$90\%$-CR & 0.0064 & 0.236 (2.19) & \textbf{0.897} & \ & 0.756 & 0.237 (0.60) & \textbf{0.898} \\
$95\%$-CR & 0.0023 & 0.260 (2.80) & \textbf{0.947} & \ & 0.740 & 0.264 (0.70) & \textbf{0.950} \\[1mm]
\midrule\\[-4mm]
\textbf{1.5-GND} & Ave.-$\beta_*$ & Ave.-Size (SE$\times 10^{-5}$) & Coverage & \ & Ave.-$\beta_*$ & Ave.-Size (SE$\times 10^{-5}$) & Coverage \\[2mm]
$80\%$-CR & 0.0181 & 0.228 (2.64) & \textbf{0.798} & \ & 0.755 & 0.230 (0.91) & \textbf{0.828} \\
$90\%$-CR & 0.0065 & 0.262 (3.55) & \textbf{0.892} & \ & 0.734 & 0.263 (0.88) & \textbf{0.914} \\
$95\%$-CR & 0.0023 & 0.284 (4.73) & \textbf{0.925} & \ & 0.716 & 0.294 (0.92) & \textbf{0.952} \\[1mm]
\bottomrule
\end{tabular} 
\end{footnotesize}
\caption{Average sizes and empirical coverages of depth-based percentile bootstrap confidence for $B = 5\, 000$ bootstrap samples and $N = 1\, 000$ simulations, using \texttt{pdMean()} and \texttt{pdDepth()}. \label{tab:1}}
\end{table}
\begin{remark}
Note that the depth-based confidence regions are equivariant under matrix congruence transformations of the sample $a \ast \bs{X} = \{a \ast X_1,\ldots, a \ast X_n \}$, with $a \in \te{GL}(d, \mathbb{C})$, in the sense that $\te{CR}_{1-\alpha}(a \ast \bs{X}) = \{ a \ast x \, :\, x \in \te{CR}_{1-\alpha}(\bs{X}) \}$. This is an immediate consequence of property \textbf{P.1} and the fact that the intrinsic mean is general linear congruence equivariant, i.e., $\mathbb{E}_{\nu}[a \ast X] = a \ast \mathbb{E}_\nu[X]$.
\end{remark}
Table \ref{tab:1} displays the empirical coverage of the percentile bootstrap confidence regions for simulated samples $X_1,\ldots,X_n \overset{\te{iid}}{\sim} \nu^p_{\te{Id}}$, with $\nu^p_{\te{Id}} \in P_2(\mathcal{M})$ a centrally symmetric distribution around the identity matrix simulated from a $p$-generalized normal distribution ($p$-GND) equivalent to the data generating processes in Figure \ref{fig:5}. The column \emph{Ave.-$\beta_*$} displays the average lower depth confidence bounds, using the notation for $\beta_*$ as in eq.(\ref{eq:5.1}). The column \emph{Ave.-Size} displays the distance of the center of the confidence ball to the furthest boundary, i.e., $\max_{\{i : D(\bar{m}_i^*, \bar{\nu}_B^*) \geq \beta_* \}} \delta_R(\bar{m}, \bar{m}_i^*)$, averaged across simulations, and the coverage is the proportion of times the identity matrix has a depth value larger or equal to the lower depth bound $\beta_*$.
\section{Analysis of multicenter clinical trial data} \label{sec:6}
\begin{figure}
  \centering
  \includegraphics[scale=0.75]{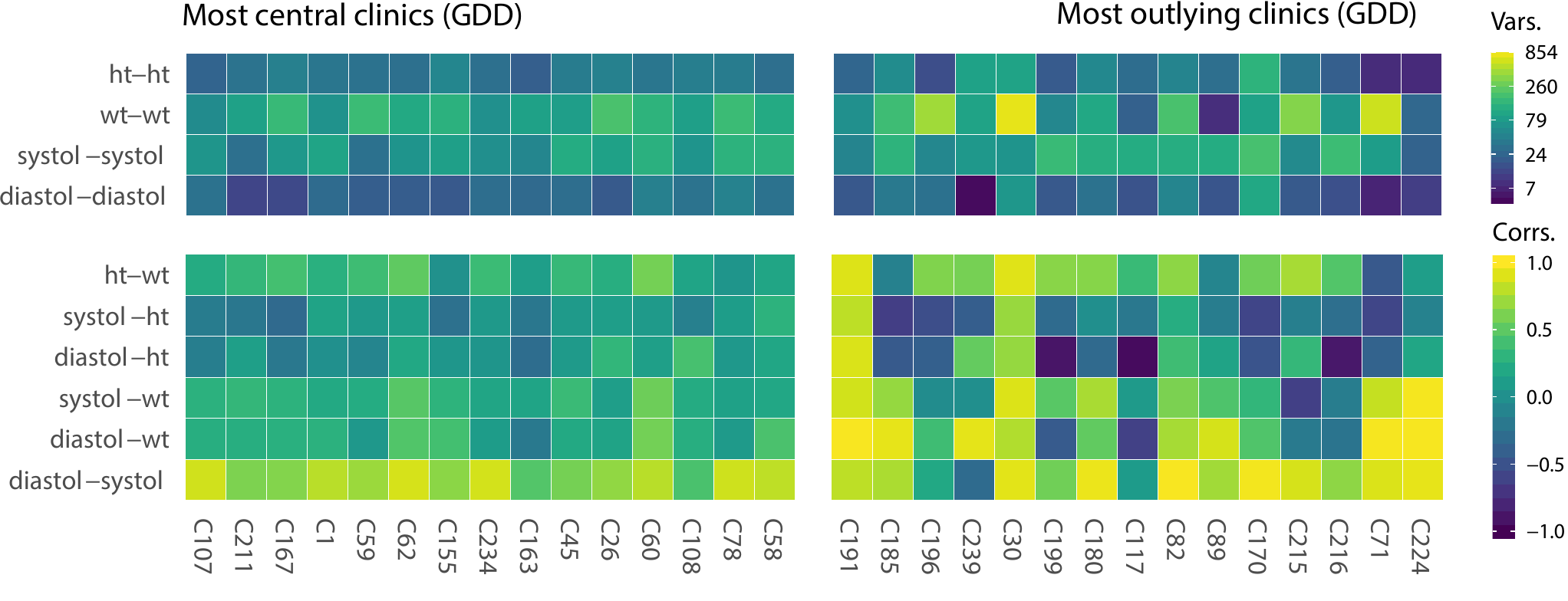}
  \caption{Most central (left) and most outlying (right) anonymized clinical centers based on the geodesic distance depth. Columns represent the clinical centers, rows represent the variances and cross-correlations. \label{fig:7}}
\end{figure}
\begin{figure}
\centering
\begin{subfigure}{0.49\linewidth}
  \centering
  \includegraphics[scale=0.75]{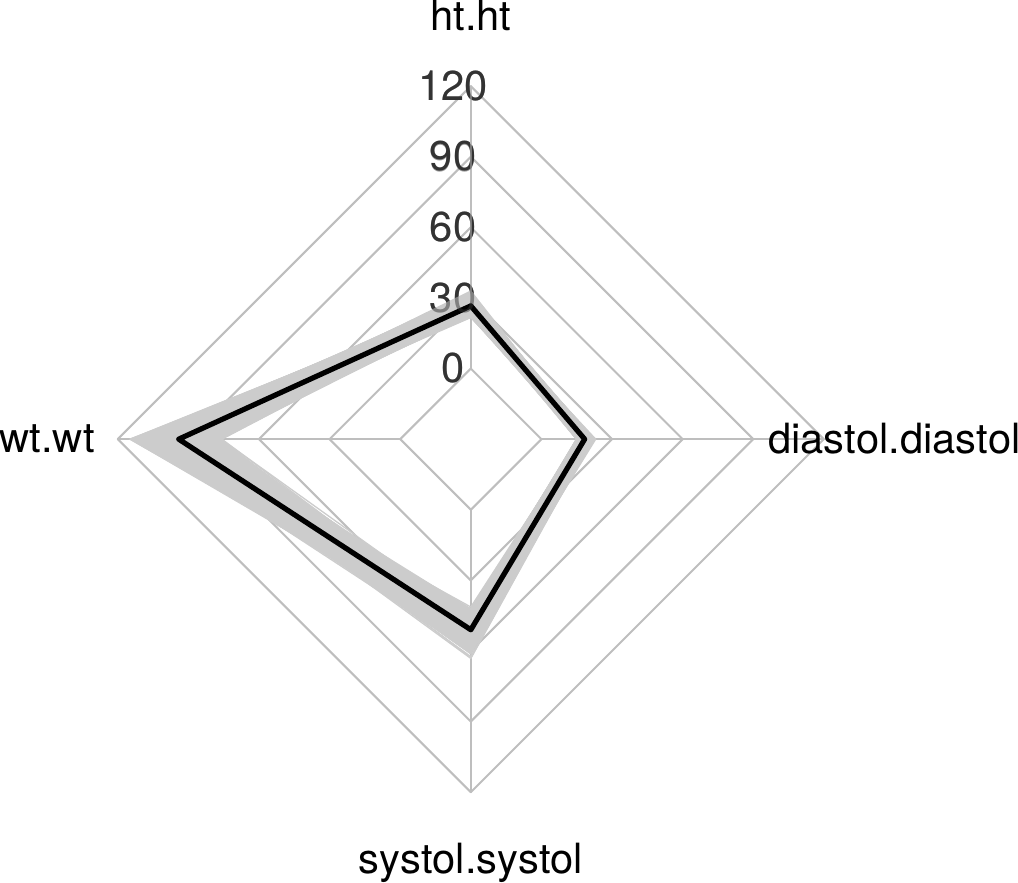}
\end{subfigure}
\begin{subfigure}{0.49\linewidth}
  \centering
  \includegraphics[scale=0.75]{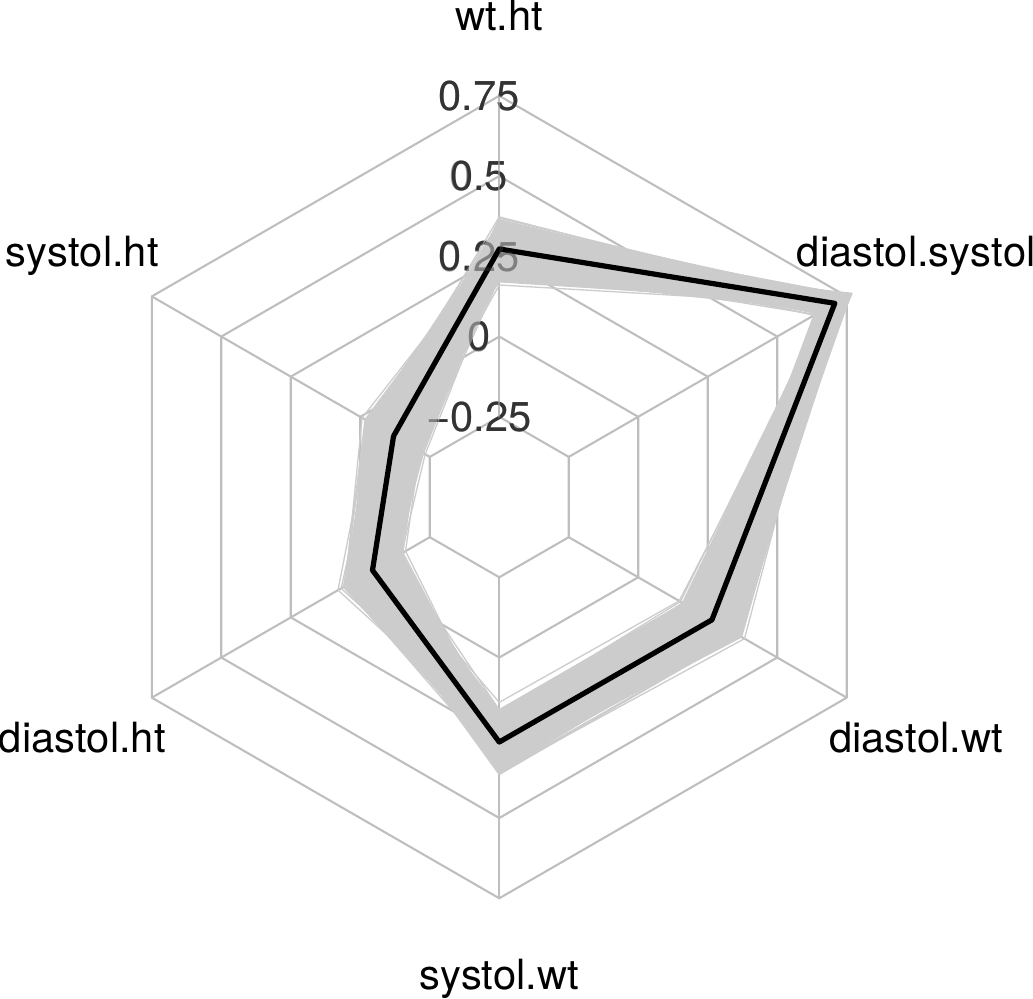}
\end{subfigure}
\caption{Radarcharts of the intrinsic sample mean (black) covariance matrix across clinical centers and 95-$\%$ intrinsic geodesic distance depth bootstrapped confidence region (grey) based on $20\, 000$ bootstrapped samples. \label{fig:8}}
\end{figure}

The intrinsic data depth functions provide a fast and intuitive procedure to explore samples of covariance matrices by identifying most central or most outlying covariance matrices, based on the Riemannian geometry of the space. This is illustrated by the exploratory analysis of a collection of sample covariance matrices obtained from 246 clinical centers (\texttt{C1}-\texttt{C246}), which have been anonymized for reasons of confidentiality. For each clinical center, medical analysts have recorded the height (\texttt{ht}), weight (\texttt{wt}), systolic blood pressure (\texttt{systol}) and diastolic blood pressure (\texttt{diastol}) for a number of clinical center patients. As part of a broader analysis, we explore the variability among clinical centers in terms of the second-order behavior, i.e., the variance-covariance structure, of the measured variables. On the one hand, we wish to identify outlying clinical centers to be flagged for further inspection or removal in subsequent data analysis. On the other hand, we are interested in the average or mean behavior of the sample covariance matrices across clinical centers.\\
Addressing the first objective, the left image in Figure \ref{fig:7} displays the 15 most central depth-ranked clinical centers (from left to right, with most central clinic \texttt{C107}) based on the geodesic distance depth applied to the collection of 246 $(4 \times 4)$-dimensional symmetric positive definite covariance matrices. The bottom rows display the six symmetric cross-correlations \texttt{ht-wt}, \texttt{systol-wt}, \texttt{diastol-ht}, \texttt{systol-wt}, \texttt{diastol-wt} and \texttt{diastol-systol}. In addition, the top rows display the four variances \texttt{ht-ht}, \texttt{wt-wt}, \texttt{systol-systol} and \texttt{diastol-diastol}, providing information about the scale of the covariance matrices. The right image in Figure \ref{fig:7} displays the 15 most outlying depth-ranked clinical centers (from right to left, with most outlying clinic \texttt{C224}) based on the geodesic distance depth in the same fashion. The center-to-outward orderings obtained via the intrinsic zonoid depth are comparable and can be found in the supplementary material. We point out that the data depth functions capture clinical centers that are outlying primarily in terms of the correlation- or covariance-structure, (e.g., center \texttt{C191}), primarily in terms of the variance-structure, (e.g., center \texttt{C170}), or both, (e.g., center \texttt{C71}). Regarding the second objective, to assess the average behavior across covariance matrices, we display in Figure \ref{fig:8} the intrinsic sample mean of the set of 246 sample covariance matrices across clinical centers, including a 95-$\%$ intrinsic geodesic distance depth percentile bootstrap confidence region. Here, the left-hand image displays the four variances and the right-hand image displays the six cross-correlations analogous to the decomposition in Figure \ref{fig:7}. The grey confidence region displays the bootstrapped sample means contained in the confidence region $\te{CR}_{0.95}(\bs{X})$. In particular, a covariance matrix $y \in \mathbb{P}_{4 \times 4}$ is included in the confidence region $\te{CR}_{0.95}(\bs{X})$ if and only if $\te{GDD}(y, \bar{\nu}^*_B) \geq \beta_*$, where $\bar{\nu}^*_B$ is the empirical distribution of the bootstrapped sample means and $\beta_*$ denotes the lower depth-bound as in Section \ref{sec:5}.
\section{Concluding remarks}
In this paper, we studied intrinsic data depth measures acting on the Riemannian manifold of symmetric or Hermitian PD matrices. The primary focus of this work is on the Riemannian manifold equipped with the affine-invariant metric, as this is the only metric that is invariant under congruence transformation of the data as described in property \textbf{P.1} in Section \ref{sec:3.1}. However, the construction of the depth functions does not fundamentally rely on the affine-invariant metric and the equivalent notions of properties \textbf{P.2} to \textbf{P.6} are expected to hold for any Riemannian metric that constitutes a geodesically complete manifold, such as the Log-Euclidean metric as discussed in \cite{A06} among others. For each of the proposed intrinsic depth functions, (including the intrinsic spatial depth), the sample data depth values are straightforward to compute and remain computationally efficient also for relatively high-dimensional matrices, with implementations directly available in the R-package \texttt{pdSpecEst}, \cite{C17}. As such, the data depths serve as an easy-to-use data exploration tool, but also provide a practical framework for inference in the context of random samples of HPD matrices, as illustrated in Section \ref{sec:5} through the construction of depth-based confidence regions. \\
Additional material available in the package \texttt{pdSpecEst} includes implementations of several intrinsic rank-based hypothesis tests, replacing the ordinary ranks by the depth-induced ranks analogous to \cite{LS93}, \cite{CS12}, or \cite[Chapter 5]{M02} for samples of Euclidean vectors. Another interesting application of the intrinsic data depth is depth-based classification or clustering for groups or samples of covariance matrices analogous to e.g., \cite{L12}. To conclude, Hermitian or symmetric positive definite matrices play an important role in many different fields of statistical research, see \cite{PFA05}, and it is of interest to apply the intrinsic data depths in other contexts than demonstrated in this paper. For instance, applied to diffusion tensor imaging, the depth functions show potential for fast detection of anomalies or artifacts in large collections of SPD diffusion tensors.

\section*{Acknowledgements}
The authors gratefully acknowledge the financial support from the following agencies and projects: the Belgian Fund for Scientific Research FRIA/FRS-FNRS (J. Chau), the contract ‘Projet d’Actions de Recherche Concert\'ees’ (ARC) No. 12/17-045 of the ‘Communaut\'e fran\c{c}aise de Belgique’ (J. Chau and R. von Sachs), IAP research network P7/06 of the Belgian government (R. von Sachs), the US National Science Foundation and KAUST (H. Ombao). We thank Lieven Desmet and the SMCS/UCL for providing access to the anonymized clinical trial data. Computational resources have been provided by the supercomputing facilities of the CISM/UCL and the C\'ECI funded by the FRS-FNRS under convention 2.5020.11.


\begin{small}
\bibliography{Main}
\end{small}

\section{Appendix I: Proofs}
\subsection{Proof of Theorem \ref{thrm:3.1}}
\begin{proof}
\textbf{P.1} This is a direct consequence of the claim that the following two events are equivalent:
\begin{eqnarray} \label{supp_eq:1.1}
\{ \bs{0}_{d \times d} \in D_{\alpha}(\zeta_y) \} & \Leftrightarrow & \{ \bs{0}_{d \times d} \in D_{\alpha}(\zeta_{a, y}) \}, \quad 0 \leq \alpha \leq 1,
\end{eqnarray}
with $\zeta_y$ the probability measure of $\Log_y(X)$ and $\zeta_{a,y}$ the probability measure of $\Log_{a \ast y}(a \ast X)$, where $X$ has probability measure $\nu$. Here, the Euclidean zonoid trimmed region $D_{\alpha}(\zeta_y)$ is represented as a set of $(d \times d)$-dimensional Hermitian matrices, instead of an equivalent set of $d^2$-dimensional real basis component vectors, as in Section \ref{sec:3.2}, and $\bs{0}_{d \times d}$ is the zero matrix. For $\alpha = 0$, the equivalence in eq.(\ref{supp_eq:1.1}) is true by definition, since $D_0(\zeta_y) = D_0(\zeta_{a,y}) = \mathbb{R}^{d \times d}$.\\[3mm]
Suppose that $\bs{0}_{d \times d} \in D_{\alpha}(\zeta_y)$ for some $0 < \alpha \leq 1$. Noting that $T_y(\mathcal{M})$ can be identified by the real vector space of Hermitian matrices $\mathcal{H}$ for each $y \in \mathcal{M}$, by definition of the zonoid $\alpha$-trimmed region, there exists a measurable function $\tilde{g} : \mathcal{H} \to [0, \frac{1}{\alpha}]$, such that,
\begin{eqnarray*}
\int_{\mathcal{H}} \tilde{g}(z)\ \zeta_y(dz)\ =\ 1, & \quad & \int_{\mathcal{H}} z \tilde{g}(z)\ \zeta_y(dz)\ =\ \bs{0}_{d \times d}.
\end{eqnarray*}
It is straightforward to verify that for each $a \in \te{GL}(d, \mathbb{C})$ and $x,y \in \mathcal{M}$, $\Log_{a \ast y}(a \ast X) = a \ast \Log_y(x)$. Define $g(z) = \tilde{g}(a^{-1} \ast z)$, then $g : \mathcal{H} \to [0, \frac{1}{\alpha}]$ is a measurable function such that,
\begin{eqnarray*}
\int_{\mathcal{H}} g(z)\ \zeta_{a,y}(dz)\ =\ \int_{\mathcal{H}} g(a \ast z)\ \zeta_{y}(dz)\ =\ \int_{\mathcal{H}} \tilde{g}(z)\ \zeta_y(dz)\ =\ 1,
\end{eqnarray*}
and,
\begin{eqnarray*}
\int_{\mathcal{H}} z g(z)\ \zeta_{a,y}(dz) \ =\ \int_{\mathcal{H}} (a \ast z) g(a \ast z)\ \zeta_{y}(dz) &=& \int_{\mathcal{H}} (a \ast z) \tilde{g}(z)\ \zeta_{y}(dz) \nn
&=& a \ast \left( \int_{\mathcal{H}} z \tilde{g}(z)\ \zeta_{y}(dz) \right) \nn
&=& a \ast \bs{0}_{d \times d}\ =\ \bs{0}_{d \times d}.
\end{eqnarray*}
Therefore $\bs{0}_{d \times d} \in D_{\alpha}(\zeta_{a,y})$. The other direction follows by a similar argument, using that $a \neq \bs{0}_{d \times d}$.\\[3mm]
\textbf{P.2} The zonoid trimmed region $D_1(\zeta_y)$ contains the single point $\bs{E}_\nu[\Log_y(X)]$ by construction. The deepest point $y \in \mathcal{M}$ is therefore characterized by the point that satisfies $\bs{E}_{\nu}[\Log_y(X)] = \bs{0}_{d \times d}$. By eq.(\ref{eq:2.4}) in the main document, on the Riemannian manifold $\mathcal{M}$ with $\nu \in P_2(\mathcal{M})$, this point is the uniquely existing geometric expectation of the distribution $\nu$.\\[3mm]
\textbf{P.3} Using the equivalent definition $\te{ZD}_{\mathcal{M}}(\nu, y) = \sup\{ \alpha : y \in D_{\alpha}^{\mathcal{M}}(\nu) \}$, by construction $D_{\alpha}^{\mathcal{M}}(\nu)$ is a geodesically convex set that contains the geometric mean $\mu := \mathbb{E}_{\nu}[X]$ for each $\alpha \in [0,1]$. Also, $D_{\alpha_1}^{\mathcal{M}}(\nu) \subseteq D_{\alpha_2}^{\mathcal{M}}(\nu)$ for each $1 \geq \alpha_1 \geq \alpha_2 \geq 0$. Combining the above arguments, it follows that a geodesic curve $\Exp_{\mu}(th)$, with $t \geq 0$ increasing, has monotone non-increasing depth as it moves further away from the center $\mu$.\\[3mm]
\textbf{P.4} With the same notation as above, for $\alpha \in (0,1]$ we claim that the sets $D_{\alpha}^{\mathcal{M}}(\nu)$ are closed and bounded, and therefore also compact by the Hopf-Rinow theorem. The fact that the sets are closed follows directly from the definition of $D_{\alpha}^{\mathcal{M}}(\nu)$. The fact that they are bounded is seen as follows; for $\alpha > 0$, by construction $D_{\alpha}^{\mathcal{M}}(\nu) \subset \mathcal{M}$. Therefore, if $y \in D_{\alpha}^{\mathcal{M}}(\nu)$, necessarily $\delta_R(\te{Id}, y) < \infty$, which follows by the fact that both $\te{Id}$ and $y$ are elements of $\mathcal{M}$, combined with \cite[Theorem 6.1.6]{B09}. Let $(y_n)_{n \in \mathbb{N}}$ be an unbounded sequence, such that $\Vert \Log(y_n) \Vert_F \to \infty$ as $n \to \infty$. The divergence $\Vert \Log(y_n) \Vert_F \to \infty$ implies in particular also that $\delta_R(\te{Id}, y_n) \to \infty$, which violates the boundedness (or compactness) of $D_{\alpha}^{\mathcal{M}}(\nu)$ for $a \in (0,1]$, and therefore we must have $\lim_{n \to \infty} \te{ZD}_{\mathcal{M}}(\nu, y_n) = \lim_{n \to \infty} \sup\{ \alpha : y_n \in D^{\mathcal{M}}_{\alpha}(\nu) \} = 0$.
\end{proof}
\subsection{Proof of Lemma \ref{lem:3.2}}
\begin{proof}
By definition of the intrinsic zonoid trimmed regions $D_{\alpha}^{\mathcal{M}}(\nu) = \{ y \in \mathcal{M} : \bs{0}_{d \times d} \in D_{\alpha}(\zeta_y) \}$ with $D_{\alpha}(\zeta_y)$ as in eq.(\ref{supp_eq:1.1}). The distribution $\zeta_y$ has finite first moment with respect to the Riemannian metric in $T_y(\mathcal{M})$, since
\begin{eqnarray*}
\int_{T_{y}(\mathcal{M})} \Vert z \Vert_y\: \zeta_y(dz) &=& \int_{\mathcal{M}} \Vert \Log_y(x) \Vert_y\: \nu(dx) \nn
&=& \int_{\mathcal{M}} \delta_R(y, x)\: \nu(dx)\ <\ \infty,
\end{eqnarray*}
using eq.(\ref{eq:2.3}) in the main document and the fact that $\nu \in P_2(\mathcal{M}) \subset P_1(\mathcal{M})$. By \cite[Theorem 3.13]{M02} for a probability measure $\zeta_y$ defined on $T_y(\mathcal{M}) \cong \mathbb{R}^{d^2}$ with finite first moments, 
\begin{eqnarray*}
\bigcup_{\alpha > 0} D_\alpha(\zeta_y) &=& \te{conv}_{T_y(\mathcal{M})}(\zeta_y),
\end{eqnarray*}
where $\te{conv}_{T_y(\mathcal{M})}(\zeta_y)$ denotes the convex hull of the support of $\zeta_y$ in $T_y(\mathcal{M}) \cong \mathbb{R}^{d^2}$, based on the Riemannian metric on $T_y(\mathcal{M})$, i.e., a rescaled Euclidean metric. Using the above result, we write out:
\begin{eqnarray*}
\bigcup_{\alpha > 0} D_{\alpha}^{\mathcal{M}}(\nu) &=& \bigcup_{\alpha > 0} \{y \in \mathcal{M} : \bs{0}_{d \times d} \in D_{\alpha}(\zeta_y) \} \nn
&=& \left\{ y \in \mathcal{M} :  \bs{0}_{d \times d} \in \cup_{\alpha > 0} D_\alpha(\zeta_y) \right\} \nn
&=& \left\{ y \in \mathcal{M} : \bs{0}_{d \times d} \in \te{conv}_{T_y(\mathcal{M})}(\zeta_y) \right\} \nn
&=& \Bigg\{ y \in \mathcal{M} : \exists\: g : \te{supp}(\nu) \to [0,1]\ \te{measurable},\ \te{s.t.}  \nn
&& \hspace{2cm} \int_{\te{supp}(\nu)} \Log_y(x)g(x)\: \lambda(dx) = \bs{0}_{d \times d}\ \te{and } \int_{\te{supp}(\nu)} g(x)\: \lambda(dx) = 1  \Bigg\} \nn 
&=& \te{conv}(\nu),
\end{eqnarray*}
where the last step follows by definition $\te{conv}(\nu)$ as the geodesic convex hull of the support of $\nu$ on the manifold.
\end{proof}
\subsection{Proof of Theorem \ref{thrm:3.3}}
\subsubsection{Continuity in $y$ (P.5)}
\begin{proof} We argue that the map $y \mapsto \te{ZD}_{\mathcal{M}}(\nu, y)$ is both upper- and lower-semicontinuous for $y \in \te{conv}(\nu)$. \\[3mm]
\textbf{Upper-semicontinuity:} the map is upper-semincontinuous if and only if for each $\alpha \in [0,1]$ the sets $\{ y \in \te{conv}(\nu) : \te{ZD}_{\mathcal{M}}(\nu, y) < \alpha \}$ are open in $\te{conv}(\nu)$ or equivalently the sets $\{ y \in \te{conv}(\nu) : \te{ZD}_{\mathcal{M}}(\nu, y) \geq \alpha \}$ are closed in $\te{conv}(\nu)$. If $\alpha = 0$, $\{ y \in \te{conv}(\nu) : \te{ZD}_{\mathcal{M}}(\nu, y) \geq \alpha \} = \te{conv}(\nu)$, and $\te{conv}(\nu)$ is closed in itself. If $\alpha > 0$, note that we can rewrite $\{y \in \te{conv}(\nu) : \te{ZD}_{\mathcal{M}}(\nu, y) \geq \alpha \} = \{ y \in \te{conv}(\nu) : y \in D^{\mathcal{M}}_\alpha(\nu) \}$, since on the one hand, if $y \in D^{\mathcal{M}}_{\alpha}(\nu)$, then $\te{ZD}_{\mathcal{M}}(\nu, y) = \sup\{ \beta : y \in D^{\mathcal{M}}_{\beta}(\nu) \} \geq \alpha$, and on the other hand, if $\te{ZD}_{\mathcal{M}}(\nu, y) = \beta \geq \alpha$, then $y \in D^{\mathcal{M}}_{\beta}(\nu) \subseteq D^{\mathcal{M}}_{\alpha}(\nu)$ by nestedness of the intrinsic zonoid trimmed regions. For each $\alpha > 0$, by construction $D_{\alpha}^{\mathcal{M}}(\nu)$ is closed, therefore $\{ y \in \te{conv}(\nu) : \te{ZD}_{\mathcal{M}}(\nu, y) \geq \alpha \}$ is also closed. \\[3mm]
\textbf{Lower-semicontinuity:} the map is lower-semicontinuous if and only if for each $\alpha \in [0,1]$ the sets $\{ y \in \te{conv}(\nu) : \te{ZD}_{\mathcal{M}}(\nu, y) \leq \alpha \}$ are closed in $\te{conv}(\nu)$ or equivalently the sets $\{ y \in \te{conv}(\nu) : \te{ZD}_{\mathcal{M}}(\nu, y) > \alpha \}$ are open in $\te{conv}(\nu)$. If $\alpha = 1$, $\{ y \in \te{conv}(\nu) : \te{ZD}_{\mathcal{M}}(\nu, y) > \alpha \} = \emptyset$, and the empty set is open in $\te{conv}(\nu)$. If $\alpha = 0$, $\{ y \in \te{conv}(\nu) : \te{ZD}_{\mathcal{M}}(\nu, y) > \alpha \} = \te{conv}(\nu)$ by Lemma \ref{lem:3.2}, and $\te{conv}(\nu)$ is open in itself. If $0 < \alpha < 1$, note that we can rewrite $\{y \in \te{conv}(\nu) : \te{ZD}_{\mathcal{M}}(\nu, y) > \alpha \} = \{y \in \te{conv}(\nu) : y \in D^{\mathcal{M}}_{\alpha+}(\nu) \}$, where,
\begin{small}
\begin{eqnarray*}
D_{\alpha+}^{\mathcal{M}}(\nu)\ :=\ \Bigg\{ y \in \mathcal{M} : y = \Exp_{y}\left(\int_{\mathcal{M}} \Log_{y}(x) g(x)\: \nu(dx) \right),
g:\mathcal{M} \to [0, 1/\alpha),  \int_{\mathcal{M}} g(x)\: \nu(dx) = 1 \Bigg\},
\end{eqnarray*}
\end{small}
with $g$ measurable.
To see that the set-equivalence is true: on the one hand, if $y \in D^{\mathcal{M}}_{\alpha+}(\nu)$, then $\te{ZD}_{\mathcal{M}}(\nu, y) = \sup\{ \beta : y \in D_{\beta}^{\mathcal{M}}(\nu) \} > \alpha$, since $[0, 1/\alpha) \subset [0, 1/\alpha]$. On the other hand, if $\te{ZD}_{\mathcal{M}}(\nu, y) = \beta > \alpha$, take $\epsilon > 0$ sufficiently small such that $\beta > \beta - \epsilon > \alpha$, then $[0, \frac{1}{\beta}] \subset [0, \frac{1}{\beta - \epsilon}) \subset [0, \frac{1}{\alpha})$. As a consequence, $y \in D^{\mathcal{M}}_{\beta}(\nu) \subseteq D^{\mathcal{M}}_{\alpha+}(\nu)$ by nestedness of the intrinsic zonoid trimmed regions. For $0 < \alpha < 1$, distinguish between two cases: (i) $D^{\mathcal{M}}_{\alpha+}(\nu) = \te{conv}(\nu)$, then the set is open as $\te{conv}(\nu)$ is open in itself, (ii) $D^{\mathcal{M}}_{\alpha+}(\nu) \subset \te{conv}(\nu)$. In this case, writing $r\partial D^{\mathcal{M}}_{\alpha+}(\nu)$ for the relative boundary of the geodesic convex set $D_{\alpha+}^{\mathcal{M}}(\nu)$ in $\te{conv}(\nu)$, we note that $r\partial D^{\mathcal{M}}_{\alpha+}(\nu) = r\partial D^{\mathcal{M}}_{\alpha}(\nu)$. Here, the relative boundary of $D_{\alpha}^{\mathcal{M}}(\nu)$ is characterized by those points in $D_{\alpha}^{\mathcal{M}}(\nu)$ for which the weighting function $g$ attains the maximum value $\frac{1}{\alpha}$. Since $D^{\mathcal{M}}_{\alpha+}(\nu) \cap r\partial D^{\mathcal{M}}_{\alpha+}(\nu) = D^{\mathcal{M}}_{\alpha+}(\nu) \cap r\partial D^{\mathcal{M}}_{\alpha}(\nu) = \emptyset$, it follows that $D^{\mathcal{M}}_{\alpha+}(\nu)$ is open. By combining the above arguments, we conclude that $\{ y \in \te{conv}(\nu) : \te{ZD}_{\mathcal{M}}(\nu, y) > \alpha \}$ is open for each $\alpha \in [0,1]$.\\[3mm]
Since the map $y \mapsto \te{ZD}_{\mathcal{M}}(\nu, y)$ is both upper- and lower-semicontinuous on $\te{conv}(\nu)$ it is also continuous on $\te{conv}(\nu)$.
\end{proof}
\subsubsection{Uniform continuity in $\nu$ (P.6)}
\begin{proof}
\textbf{Pointwise convergence of depths:} first, we show pointwise convergence of $\te{ZD}_{\mathcal{M}}(\nu_n, y)$ to $\te{ZD}_{\mathcal{M}}(\nu, y)$ for each $y \in \te{rint}(\te{conv}(\nu))$, where $\te{rint}(\te{conv}(\nu))$ denotes the relative interior of the geodesic convex set $\te{conv}(\nu)$. We note that $y \in \te{rint}(\te{conv}(\nu))$ if and only if $\bs{0}_{d \times d} \in \te{rint}(\te{conv}_{T_y(\mathcal{M})}(\zeta_y))$, where $\te{conv}_{T_y(\mathcal{M})}(\zeta_y)$ is the convex hull of the support of $\zeta_y$ in $T_y(\mathcal{M})$ as in the proof of Lemma \ref{lem:3.2}. This is seen as follows: by Lemma \ref{lem:3.2}, $y \in \te{conv}(\nu)$ if and only if $\exists\: \alpha > 0$, such that $y \in D_{\alpha}^{\mathcal{M}}(\nu)$, but this is equivalent to $\bs{0}_{d \times d} \in D_{\alpha}(\zeta_y)$ which holds if and only if $\bs{0}_{d \times d} \in \te{conv}_{T_y(\mathcal{M})}(\zeta_y)$ by \cite[Theorem 3.13]{M02}. Since the sets $\{ y : y \in \te{conv}(\nu) \}$ and $\{ y : \bs{0}_{d \times d} \in \te{conv}_{T_y(\mathcal{M})}(\zeta_y) \}$ are equivalent their relative interiors are equivalent as well. By Definition \ref{def:3.1}, $\te{ZD}_{\mathcal{M}}(\nu_n, y) = \te{ZD}_{\mathbb{R}^{d^2}}(\zeta^n_y, \vec{0})$, where $\zeta^n_y$ is the distribution of $\Log_y(X)$ as a $d^2$-dimensional real basis component vector, with $X \sim \nu_n$, such that $\zeta_y^n \overset{w}{\to} \zeta_y$. Similarly, $\te{ZD}_{\mathcal{M}}(\nu, y) = \te{ZD}_{\mathbb{R}^{d^2}}(\zeta_y, \vec{0})$. By the same argument as in the proof of Lemma \ref{lem:3.2}, we know that $\zeta^n_y, \zeta_y \in P_1(T_y(\mathcal{M}))$ for each $n \in \mathbb{N}$, where $P_1(T_y(\mathcal{M}))$ denotes the set of probability measures on $T_y(\mathcal{M})$ with finite first moment, i.e., if $\zeta \in P_1(T_y(\mathcal{M}))$ then $\int_{T_y(\mathcal{M})} \Vert z \Vert_y\: d\zeta_y(z) < \infty$. Furthermore, the sequence of measures $(\zeta^n_{y})_{n \in \mathbb{N}}$ is uniformly integrable with respect to the Riemannian metric in $T_y(\mathcal{M})$, since for any $y \in \mathcal{M}$,
\begin{eqnarray*}
\lim_{K \to \infty} \sup_{n \in \mathbb{N}} \int_{T_y(\mathcal{M})} \Vert z \Vert_y \bs{1}_{\{ \Vert z \Vert_y > K \}}\: \zeta^n_{y}(dz)  \ = \ \lim_{K \to \infty} \sup_{n \in \mathbb{N}} \int_{\mathcal{M}} \delta_R(y,x) \bs{1}_{\{ \delta_R(y, x) > K \}}\: \nu_n(dx)\ =\ 0.
\end{eqnarray*}
By \cite[Theorem 4.6]{M02}, under these conditions, for $y \in \te{rint}(\te{conv}(\nu))$ or equivalently $\bs{0}_{d \times d} \in \te{rint}(\te{conv}_{T_y(\mathcal{M})}(\zeta_y))$, it follows that, 
\begin{eqnarray} \label{supp_eq:1.2}
\te{ZD}_{\mathcal{M}}(\nu_n, y)\ =\ \te{ZD}_{\mathbb{R}^{d^2}}(\zeta^n_y, \vec{0}) & \to & \te{ZD}_{\mathbb{R}^{d^2}}(\zeta_y, \vec{0}) \ = \ \te{ZD}_{\mathcal{M}}(\nu, y), \quad \quad \te{as } n \to \infty. \quad \quad
\end{eqnarray}
\textbf{Uniform convergence of depths:} uniform depth convergence now follows from the pointwise depth convergence above by a generalized version of the proof of \cite[Theorem 4.8]{D16} for the complete metric space $(\mathcal{M}, \delta_R)$, using Lemma \ref{lem:3.2} and the fact that $\te{ZD}_{\mathcal{M}}(\nu, y)$ is a \emph{normed} geodesically convex depth, continuous in $y$ by the first part of Theorem \ref{thrm:3.3}. Since the proof is completely analogous to the proof of \cite[Theorem 4.8]{D16}, we omit the details here. Note that the only required modification is to replace the Euclidean metric space by the complete metric space $(\mathcal{M}, \delta_R)$. In particular, Euclidean open balls, convex sets and convergence are replaced by geodesic open balls, geodesic convex sets and convergence in the Riemannian distance function respectively.\\[3mm]
By the generalized proof of \cite[Theorem 4.8]{D16}, the depths $(\te{ZD}_{\mathcal{M}}(\nu_n, y_0))_{n \in \mathbb{N}}$ are continuously convergent for $y_0 \in \te{rint}(\te{conv}(\nu))$. That is, for $y_n \to y_0$ in the sense that $\delta_R(y_n, y_0) \to 0$, also $\lim_{n \to \infty} \te{ZD}_{\mathcal{M}}(\nu_n, y_n) = \te{ZD}(\nu, y_0)$. By \cite[Proposition A.1]{D16}, since $\mathcal{M}$ is a metric space, continuous convergence of the depths implies compact convergence, i.e., for every compact set $M \subseteq \te{rint}(\te{conv}(\nu))$,
\begin{eqnarray*}
\lim_{n \to \infty} \sup_{y \in M} |\te{ZD}_{\mathcal{M}}(\nu_n, y) - \te{ZD}_{\mathcal{M}}(\nu, y)| &=& 0.
\end{eqnarray*}
Consequently, by \cite[Theorem 4.4]{D16}, compact convergence implies uniform convergence, since the arguments in the proof of \cite[Theorem 4.4]{D16} continue to hold for the intrinsic zonoid depth defined on the complete metric space $\mathcal{M}$, where closed and bounded subsets are compact.
\end{proof}

\subsection{Proof of Theorem \ref{thrm:3.4} and Proposition \ref{prop:3.5}}
\begin{proof}
Properties \textbf{P.1}--\textbf{P.4} follow directly by Theorem \ref{thrm:3.1}, using the definition of the depth as the integrated pointwise zonoid depth (integrated over $t \in \mathcal{I}$). \\[3mm]
For the first part (\textbf{P.5}) of Proposition \ref{prop:3.5}: using that $\sup_{t \in \mathcal{I}} \delta_R(y_n(t), y(t)) \to 0$, by the first part of Theorem \ref{thrm:3.3}, $\te{ZD}_{\mathcal{M}}(\nu(t), y_n(t)) \to \te{ZD}_{\mathcal{M}}(\nu(t), y(t))$ uniformly over $t \in \mathcal{I}$. By definition of the integrated intrinsic zonoid depth also,
\begin{eqnarray*}
|\te{iZD}_{\mathcal{M}}(\nu, y_n) - \te{iZD}_{\mathcal{M}}(\nu, y)| \ \leq \ \int_{\mathcal{I}} |\te{ZD}_{\mathcal{M}}(\nu(t), y_n(t)) - \te{ZD}_{\mathcal{M}}(\nu(t), y(t))|\ dt \ \to \ 0,
\end{eqnarray*}
by the pointwise convergence and the fact that the depth function $\te{ZD}_{\mathcal{M}}(\cdot, \cdot) \in [0,1]$ is bounded. \\[3mm]
For the second part (\textbf{P.6}) in Proposition \ref{prop:3.5}: under the given assumptions, by the second part of Theorem \ref{thrm:3.3},
\begin{eqnarray*}
\sup_{y(t) \in \te{rint}(\te{conv}(\nu(t))} |\te{ZD}_{\mathcal{M}}(\nu_n(t), y(t)) - \te{ZD}_{\mathcal{M}}(\nu(t), y(t))| & \to & 0, \quad \te{uniformly for } t \in \mathcal{I},
\end{eqnarray*}
and similarly as above,
\begin{eqnarray*}
\lefteqn{\sup_{y \in \te{rint}(\te{conv}(\nu))} | \te{iZD}_{\mathcal{M}}(\nu_n, y) - \te{iZD}_{\mathcal{M}}(\nu, y)| \ \leq \ }\nn
&& \hspace{4cm} \sup_{y \in \te{rint}(\te{conv}(\nu))} \int_{\mathcal{I}} | \te{ZD}_{\mathcal{M}}(\nu_n(t), y(t)) - \te{ZD}_{\mathcal{M}}(\nu(t), y(t))|\ dt \ \to \ 0,
\end{eqnarray*}
using the pointwise convergence and the fact that the depth function $\te{ZD}_{\mathcal{M}}(\cdot, \cdot) \in [0,1]$ is bounded.
\end{proof}

\subsection{Proof of Theorem \ref{thrm:3.6}}
\begin{proof}
\textbf{P.1} This follows directly from the definition of the depth by the fact that the map $x \mapsto a \ast x$ with $a \in \te{GL}(d, \mathbb{C})$ is distance preserving, i.e., $\delta_R(a \ast x, a \ast y) = \delta_R(x, y)$ for each $x, y \in \mathcal{M}$.\\[3mm]
\textbf{P.2} Since $\int_{\mathcal{M}} \delta_R(y, x)\: \nu(dx) \geq 0$ and $\exp(-z)$ is strictly decreasing in $z \geq 0$, the point of maximum depth is attained at $y = \arg\min_{z \in \te{supp}(\nu)} \int_{\mathcal{M}} \delta_R(z, x)\: \nu(dx)$. By eq.(\ref{eq:2.5}) in the main document, on the Riemannian manifold $\mathcal{M}$ with $\nu \in P_1(\mathcal{M})$, this point is the uniquely existing geometric median of the distribution $\nu$.\\[3mm]
\textbf{P.3} By the proof of \cite[Theorem 1]{F09} and an application of Leibniz's integral rule, $y \mapsto \bs{E}_{\nu}[\delta_R(y, X)]$ is a (strictly) convex function, and by \textbf{P.2}\@ it attains its unique minimum at $m := \te{GM}\nu(X)$. This implies that $\bs{E}_{\nu}[\delta_R(\Exp_m(th), X)]$ is a nondecreasing function for $t \geq 0$, where $\Exp_m(th)$ is a geodesic curve emanating from $m$ with unit tangent vector $h$. As a consequence $\te{GDD}(\nu, \Exp_m(th)) = \exp\left( - \bs{E}_{\nu}[\delta_R(\Exp_m(th), X)]\right)$ is monotone non-increasing for $t \geq 0$.\\[3mm]
\textbf{P.4} Let $(y_n)_{n \in \mathbb{N}}$ be an unbounded sequence such that $\Vert \Log(y_n) \Vert_F \to \infty$ as $n \to \infty$, then also $\delta_R(y_n, x) = \Vert \Log(x^{-1/2} \ast y_n) \Vert_F \to \infty$ for each $x \in \mathcal{M}$, and as a consequence $\te{GDD}(\nu, y_n) =  \exp(- \bs{E}_\nu[\delta_R(y_n, X)]) \to 0$.
\end{proof}

\subsection{Proof of Theorem \ref{thrm:3.7}}
\subsubsection{Continuity in $y$ (P.5)}
\begin{proof}
First, suppose that $(y_n)_{n \in \mathbb{N}}$ is an unbounded sequence $\Vert \Log(y_n) \Vert_F \to \infty$ as $n \to \infty$, i.e., $y_n \to y$, where $y$ is a singular matrix. Since $\te{GDD}(\nu, y) = 0$, by \textbf{P.4} in Theorem \ref{thrm:3.6}, $\lim_{n \to \infty} \te{GDD}(\nu, y_n) = \te{GDD}(\nu, y)$. Second, suppose that $(y_n)_{n \in \mathbb{N}}$ is a bounded sequence, i.e., $\sup_{n \in \mathbb{N}} \Vert \Log(y_n) \Vert_F = \sup_{n \in \mathbb{N}} \delta_R(y_n, \te{Id}) < \infty$. Since $\nu \in P_1(\mathcal{M})$, there exists an $y_0 \in \mathcal{M}$ such that $\int_{\mathcal{M}} \delta_R(y_0, x)\: \nu(dx) < \infty$. By the triangle inequality, 
\begin{eqnarray} \label{supp_eq:1.3}
\int_{\mathcal{M}} \sup_{n \in \mathbb{N}} \delta_R(y_n, x)\: \nu(dx) & \leq & \sup_{n \in \mathbb{N}} \delta_R(y_n, \te{Id}) + \delta_R(\te{Id}, y_0) +\int_{\mathcal{M}} \delta_R(y_0, x)\: \nu(dx) \ < \infty, \quad
\end{eqnarray} 
using that $\delta_R(y_0, \te{Id}) < \infty$ as both $\te{Id}$ and $y_0$ are elements of $\mathcal{M}$, (see \cite[Theorem 6.1.6]{B09}). We show continuity directly from the definition of the geodesic distance depth. The function $z \mapsto \exp(-z)$ is continuous in $z$, also the function $z \mapsto \delta_R(z, x)$ is continuous in $z$, since $\delta_R(z, x) = \Vert \Log(x^{-1/2} \ast z) \Vert_F$ is a composition of continuous functions. Furthermore, by the dominated convergence theorem, $\lim_{n \to \infty} \int_{\mathcal{M}} \delta_R(y_n, x)\: \nu(dx) = \int_{\mathcal{M}} \lim_{n \to \infty} \delta_R(y_{n}, x)\: \nu(dx)$, since $\int_{\mathcal{M}} \sup_{n \in \mathbb{N}} \delta_R(y_n, x)\: \nu(dx) < \infty$. Combining these arguments, $\lim_{n \to \infty} \te{GDD}(\nu, y_n) = \te{GDD}(\nu, y)$. 
\end{proof}
\subsubsection{Uniform continuity in $\nu$ (P.6)}
\begin{proof}
We start by noting that the uniform integrability condition implies in particular that $\nu_n \in P_1(\mathcal{M})$ for each $n \in \mathbb{N}$. Also, since $z \mapsto \delta_R(y, z)$ is continuous in $z$, by the continuous mapping theorem $\delta_R(y, X_n) \overset{d}{\to} \delta_R(y, X)$, with $X_n \sim \nu_n$ and $X \sim \nu$, and by Vitali's convergence theorem $\int_{\mathcal{M}} \delta_R(y, x)\: \nu_n(dx) \to \int_{\mathcal{M}} \delta_R(y, x)\: \nu(dx)$ for any $y \in \mathcal{M}$. Note that the convergence implies in particular also that $\nu \in P_1(\mathcal{M})$. For two measures $\mu, \nu \in P_1(\mathcal{M})$ define their $L^1$-Wasserstein distance as: 
\begin{eqnarray*}
W_1(\mu, \nu) &=& \inf_{\gamma \in \Gamma(\mu, \nu)} \int_{\mathcal{M} \times \mathcal{M}} \delta_R(y, x)\: \gamma(dy, dx),
\end{eqnarray*}
where $\Gamma(\mu, \nu)$ denotes the collection of all probability measures on $\mathcal{M} \times \mathcal{M}$ with marginal measures $\mu$ and $\nu$. Substituting $\mu = \delta_y$, the point measure in $y$, it follows that $W_1(\delta_y, \nu) = \int_{\mathcal{M}} \delta_R(y, x)\ \nu(dx)$. Therefore, a sufficient condition to ensure uniform convergence in $y \in \mathcal{M}$ of $\int_{\mathcal{M}} \delta_R(y, x)\: \nu_n(dx)$ to $\int_{\mathcal{M}} \delta_R(y, x)\: \nu(dx)$, is $W(\nu_n, \nu) \to 0$, since
\begin{eqnarray} \label{supp_eq:1.4}
\sup_{y \in \mathcal{M}} \left|\int_{\mathcal{M}} \delta_R(y, x)\: \nu_n(dx) - \int_{\mathcal{M}} \delta_R(y, x)\: \nu(dx) \right| & = & \sup_{y \in \mathcal{M}} \left| W_1(\delta_y, \nu_n) - W_1(\delta_y, \nu) \right| \nn
& \leq & W_1(\nu_n, \nu), 
\end{eqnarray}
where the last step follows by the reverse triangle inequality for the $L^1$-Wasserstein distance. The manifold $\mathcal{M}$ is a complete separable metric space, and therefore by \cite[Theorem 6.9]{V09} a necessary and sufficient condition for $W_1(\nu_n, \nu) \to 0$ is that the sequence of probability measures $\nu_n$ converges weakly in $P_1(\mathcal{M})$ to $\nu$, i.e., (i) $\nu_n \overset{w}{\to} \nu$ and (ii) $\int_{\mathcal{M}} \delta_R(y, x)\ \nu_n(dx) \to \int_{\mathcal{M}} \delta_R(y, x)\ \nu(dx)$ for any $y \in \mathcal{M}$. Condition (i) holds by assumption, and condition (ii) has already been shown above. \\[3mm]
The function $z \to \exp(-z)$ is uniformly continuous for $z \geq 0$, therefore the uniform convergence of the geodesic distance depth follows as well since,
\begin{eqnarray*}
\sup_{y \in \mathcal{M}} |\te{GDD}(\nu_n, y) - \te{GDD}(\nu, y)| &=& \sup_{y \in \mathcal{M}}|\exp(-\bs{E}_{\nu_n}[\delta_R(y, X)]) - \exp(-\bs{E}_{\nu}[\delta_R(y,X)])| \nn
& \overset{n \to \infty}{\to} & 0.  
\end{eqnarray*}
\end{proof}

\subsection{Proof of Theorem \ref{thrm:3.8} and Proposition \ref{prop:3.9}}
\begin{proof}
Properties \textbf{P.1}--\textbf{P.4} follow directly by the pointwise depth properties in Theorem \ref{thrm:3.6}, using the definition of the depth in terms of the integrated Riemannian distance (integrated over $t \in \mathcal{I}$). \\[3mm]
For the first part (\textbf{P.5}) of Proposition \ref{prop:3.9}: using that $\sup_{t \in \mathcal{I}}(\delta_R(y_n(t), y(t)) \to 0$, by the first part of the proof in Theorem \ref{thrm:3.7} also, 
\begin{eqnarray*}
\sup_{t \in \mathcal{I}}\left| \bs{E}_{\nu(t)}[ \delta_R(y_n(t), X)] - \bs{E}_{\nu(t)}[\delta_R(y(t), X)] \right| & \overset{n \to \infty}{\to} & 0,
\end{eqnarray*}
and as a direct consequence $\lim_{n \to \infty} \int_{\mathcal{I}} \bs{E}_{\nu(t)}[\delta_R(y_n(t), X)]\: dt = \int_{\mathcal{I}} \bs{E}_{\nu(t)}[\delta_R(y(t), X)]\: dt$. Using again that $z \mapsto \exp(-z)$ is continuous in $z$, the composition converges as well and we conclude that $\lim_{n \to \infty} \te{iGDD}(\nu, y_n) = \te{iGDD}(\nu, y)$.\\[3mm]
For the second part (\textbf{P.6}) of Proposition \ref{prop:3.9}. Denote by $\xi_{n,y}(t)$ and $\xi_y(t)$ respectively the distributions of $\delta_R(y(t), X_n(t))$ and $\delta_R(y(t), X(t))$, such that $X_n(t) \sim \nu_n(t)$ and $X(t) \sim \nu(t)$. Let $\phi :\mathbb{R} \to \mathbb{R}$ be a continuous and bounded function and write $y \in \mathcal{M}$ for a curve with $y(t) \in \mathcal{M}$ for each $t \in \mathcal{I}$. Then for any curve $y \in \mathcal{M}$,
\begin{eqnarray*}
\sup_{t \in \mathcal{I}} | \bs{E}_{\xi_{n,y}(t)}[\phi(X)] - \bs{E}_{\xi_y(t)}[\phi(X)]| & = & \sup_{t \in \mathcal{I}} | \bs{E}_{\nu_n(t)}[\phi(\delta_R(y(t), X))] - \bs{E}_{\nu(t)}[\phi(\delta_R(y(t), X))] | \nn
& \overset{n \to \infty}{\to} & 0,
\end{eqnarray*}
where the last step follows by the fact that for each $t \in \mathcal{I}$ the composition $x \mapsto \phi(\delta_R(y(t), x))$ is again a continuous and bounded function, and the fact that $\nu_n(t) \overset{w}{\to} \nu(t)$ uniformly in $t$. Thus, for any curve $y \in \mathcal{M}$, the weak convergence $\xi_{n,y}(t) \overset{w}{\to} \xi_y(t)$ holds as well uniformly in $t$. By the uniform integrability of $(\nu_n(t))_{n \in \mathbb{N}}$ uniformly in $t$, combined with Vitali's convergence theorem, it follows that for each curve $y \in \mathcal{M}$,
\begin{eqnarray} \label{supp_eq:1.5}
\sup_{t \in \mathcal{I}} | \bs{E}_{\nu_n(t)}[ \delta_R(y(t), X) ] - \bs{E}_{\nu(t)}[ \delta_R(y(t), X) ]| & \to & 0, \quad \quad \te{as } n \to \infty.
\end{eqnarray}
By the same argument as in the second part of the proof of Theorem \ref{thrm:3.7}, a sufficient condition for uniform convergence in $y \in \mathcal{M}$ of $\int_{\mathcal{I}} \bs{E}_{\nu_n(t)}[\delta_R(y(t), X)]\: dt$ to $\int_{\mathcal{I}} \bs{E}_{\nu(t)}[\delta_R(y(t), X)]\: dt$ is the condition $\sup_{t \in \mathcal{I}} W_1(\nu_n(t), \nu(t)) \to 0$. Again by \cite[Theorem 6.9]{V09}, the convergence $\sup_{t \in \mathcal{I}} W_1(\nu_n(t), \nu(t)) \to 0$ is implied by the conditions (i) $\nu_n(t) \overset{w}{\to} \nu(t)$ uniformly in $t$, which holds by assumption and (ii) the convergence in eq.(\ref{supp_eq:1.5}) pointwise in $y \in \mathcal{M}$.\\[3mm]
The function $z \to \exp(-z)$ is uniformly continuous for $z \geq 0$, therefore the uniform convergence of the integrated geodesic distance depth follows as well,
\begin{eqnarray*}
\lefteqn{\sup_{y \in \mathcal{M}} |\te{iGDD}(\nu_n, y) - \te{iGDD}(\nu, y)|\ =\ }\nn
&& \hspace{0.5cm} \sup_{y \in \mathcal{M}}\left|\exp\left(-\int_{\mathcal{I}} \bs{E}_{\nu_n(t)}[\delta_R(y(t), X)]\: dt \right) - \exp\left(- \int_{\mathcal{I}} \bs{E}_{\nu(t)}[\delta_R(y(t), X)]\: dt\right) \right| \ \overset{n \to \infty}{\to} \ 0.
\end{eqnarray*}
\end{proof}

\subsection{Proof of Proposition \ref{prop:4.1}}
\begin{proof}
First, we verify that $e_2(X) \leq 1/2$.\\[3mm]
Let $Y_1 = \ldots = Y_n = p \in \mathcal{M}$ be $n$ contaminating observations, such that $\Vert \Log(p) \Vert_F \geq N$ for some $N > 0$. Denote $\nu_{n,n}$ for the empirical distribution of the contaminated sample $Z^{(n,n)} = \{X_1,\ldots,X_n\} \cup \{Y_1,\ldots,Y_n\}$. For each $x \in \{X_1,\ldots,X_n \}$, 
\begin{eqnarray*}
D(Y_1, \nu_{n,n}) & = & \exp\left(-\sum_{i=1}^n \delta_R(p, X_i)\right) \nn
& \geq & \exp\left(-\sum_{i=1}^n \delta_R(X_i, x) - \sum_{i=1}^n \delta_R(x,p)\right) \ = \ D(x, \nu_{n,n}),
\end{eqnarray*}
using the triangle inequality $\delta_R(p, X_i) \leq \delta_R(p, x) + \delta_R(x, X_i)$ for each $i=1,\ldots,n$. Since $Y_1 = \ldots = Y_n$, $D(Y_1, \nu_{n,n}) = \ldots = D(Y_n, \nu_{n,n}) \geq D(x, \nu_{n,n})$ for each $x \in \{X_1,\ldots,X_n\}$. Therefore, $\Vert \Log(Z^{(n,n)}_{[1]}) \Vert_F = \ldots = \Vert \Log(Z^{(n,n)}_{[n]} )\Vert_F = \Vert \Log(p) \Vert_F \geq N$, with $Z_{[i]}^{(n,n)}$ the $i$-th depth ranked observation in the sample $Z^{(n,n)}$. As we can choose $p \in \mathcal{M}$, such that $\Vert \Log(p) \Vert_F \geq N$ for $N$ arbitrarily large, $\Vert \Log(Z^{(n,n)}_{[i]}) \Vert_F$ with $1 \leq i \leq n$ can be made arbitrarily large by adding $n$ contaminating observations. This implies that $\epsilon_2(X) \leq n/(2n) = 1/2$.\\[3mm]
Second, we verify that $\epsilon_2(X) \geq 1/2$.\\[3mm]
Consider the contaminated sample $Z^{(n,m)} = \{X_1,\ldots,X_n\} \cup \{Y_1,\ldots,Y_m\}$, with $m < n$. If we can show that $D(y, \nu_{n,m}) < D(x, \nu_{n,m})$ for each $y \in \{Y_1,\ldots,Y_m\}$ and each $x \in \{X_1,\ldots,X_n\}$. Then $\forall\, i \in\{1,\ldots,n\}$, $\exists j \in \{1,\ldots,n\}$, such that $Z^{(n,m)}_{[i]} = X_j$ and consequently $\max_i \Vert \Log(Z_{[i]}^{(n,m)}) \Vert_F \leq M$, denoting $M := \max_i \Vert \Log(X_i)\Vert_F$. The latter implies that it takes at least $m \geq n$ contaminating observations to make $\Vert \Log(Z^{(n,m)}_{[i]}) \Vert_F$ arbitrarily large for $1 \leq i \leq n$, i.e., $\epsilon_2(X) \geq 1/2$. It remains to show that $D(y, \nu_{n,m}) < D(x, \nu_{n,m})$ for each $y \in \{Y_1,\ldots,Y_m\}$ and each $x \in \{X_1,\ldots,X_n\}$.\\[3mm]
Let $y \in \{Y_1,\ldots,Y_n\}$ and $x \in \{X_1,\ldots,X_n\}$ arbitrary, then:
\begin{eqnarray} \label{eq:7.6}
D(y, \nu_{n,m}) < D(x, \nu_{n,m}) &\ \Leftrightarrow \ & \sum_{i=1}^m \delta_R(y, Y_i) + \sum_{i=1}^n \delta_R(y, X_i) > \sum_{i=1}^m \delta_R(x, Y_i) + \sum_{i=1}^n \delta_R(x, X_i).\nn
\end{eqnarray}
Let us denote $R := \max_i \delta_R(x, X_i)$, $B := \{ p \in \mathcal{M}\, :\, \delta_R(p,x) \leq 2R \}$ and $\rho = \inf_{p \in B} \delta_R(p, y)$.\\
First, by the triangle inequality $\delta_R(x, y) \leq 2R + \rho$. Therefore, by the reverse triangle inequality, $\forall\, i=1,\ldots,m$,
\begin{eqnarray} \label{eq:7.7}
\delta_R(y, Y_i) & \geq & \delta_R(x, Y_i) - \delta_R(x, y) \nn
& \geq & \delta_R(x, Y_i) - (2R + \rho).
\end{eqnarray} \label{eq:7.8}
Also, by definition of $R$ and $\rho$, $\forall\, i=1,\ldots,n$, 
\begin{eqnarray}
\delta_R(y, X_i) \ \geq \ R + \rho \ \geq \ \delta_R(x, X_i) + \rho.
\end{eqnarray}
Without loss of generality, assume that $\min_i \Vert \Log(Y_i) \Vert_F \geq N$, where $N \geq 2(n+1)R + M$. Denoting $\te{Id}$ for the identity matrix, it follows that,
\begin{eqnarray} \label{eq:7.9}
\rho \ = \ \inf_{p \in B} \delta_R(y, p) & \geq & \delta_R(y, \te{Id}) - \sup_{p \in B} \delta_R(p, \te{Id}) \nn
& \geq & N - \sup_{p \in B} (\delta_R(p, x) + \delta_R(x, \te{Id})) \nn
& \geq & 2(n+1)R + M - (2R + M) \ = \ 2nR.
\end{eqnarray}
Here, we used two triangle inequalities and the fact that $\Vert \Log(z) \Vert_F = \delta_R(z, \te{Id})$ by definition of the Riemannian distance. Combining eq.(\ref{eq:7.7}-\ref{eq:7.9}) above yields:
\begin{eqnarray*}
\sum_{i=1}^m \delta_R(y, Y_i) + \sum_{i=1}^n \delta_R(y, X_i) & \geq & \sum_{i=1}^m (\delta_R(x, Y_i) + (2R + \rho)) + \sum_{i=1}^n (\delta_R(x, X_i) + \rho) \nn
& \geq & -2mR + (n-m)\rho + \sum_{i=1}^m \delta_R(x, Y_i) + \sum_{i=1}^n \delta_R(x, X_i) \nn
& > & \sum_{i=1}^m \delta_R(x, Y_i) + \sum_{i=1}^n \delta_R(x, X_i),
\end{eqnarray*}
where we used that $-2mR + (n-m)\rho > -2nR + \rho \geq 0$ by the fact that $m < n$ and $\rho \geq 2nR$ by eq.(\ref{eq:7.9}). Returning to eq.(\ref{eq:7.6}), it follows that $D(y, \nu_{n,m}) < D(x, \nu_{n,m})$. As this result holds for any $y \in \{Y_1,\ldots,Y_m\}$ and $x \in \{X_1,\ldots,X_n\}$, we conclude that $\epsilon_2(X) \geq 1/2$. Since also $\epsilon_2(X) \leq 1/2$, it follows that $\epsilon_2(X) = 1/2$, which finishes the proof.
\end{proof}

\section{Appendix II: Additional figures}
\begin{figure}[ht]
  \centering
  \includegraphics[scale=0.75]{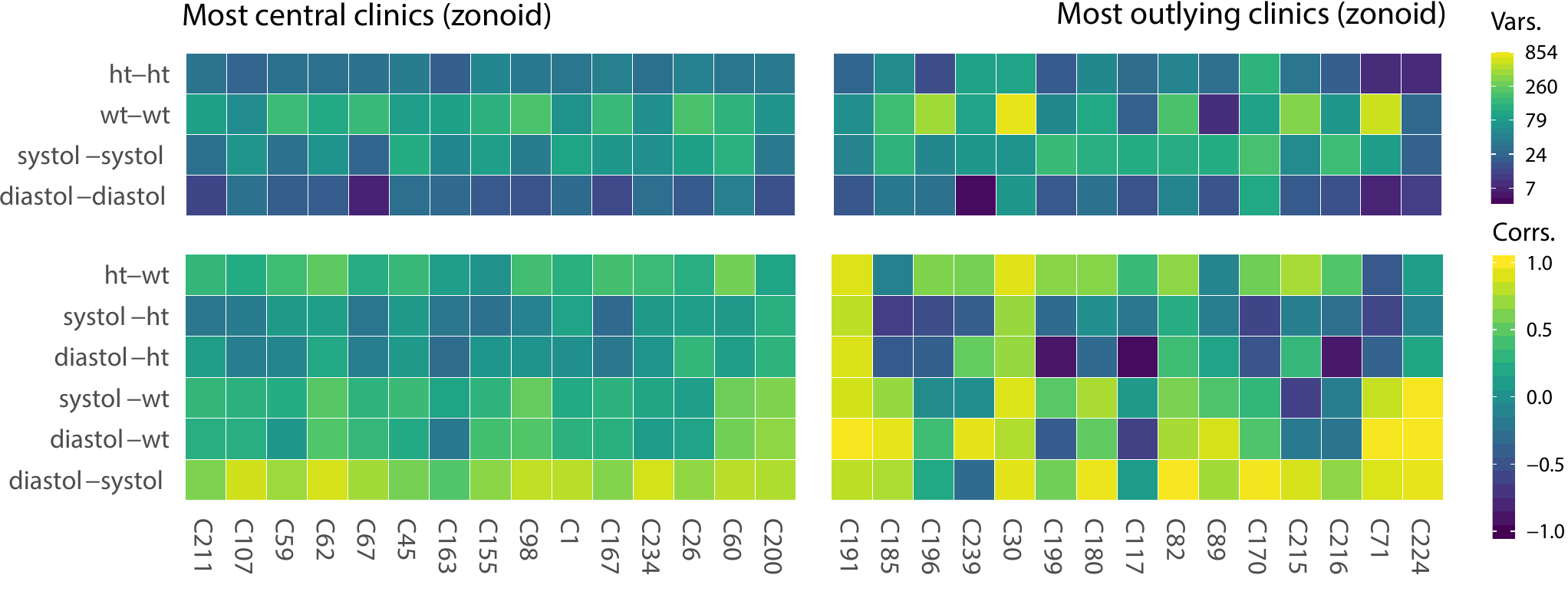}
  \caption{Most central (left) and most outlying (right) clinical centers based on the intrinsic zonoid depth analogous to Figure \ref{fig:7} in the main document for the GDD. Columns represent the clinical centers, rows represent the variances and cross-correlations. Note that we break ties in the center-to-outward zonoid depth-ranks by assigning the lowest rank to the clinical center with smallest Riemannian distance to the intrinsic sample mean across clinical centers.}
  \end{figure}
  \begin{figure}[ht]
	\centering
\begin{subfigure}{0.49\linewidth}
  \centering
  \includegraphics[scale=0.65]{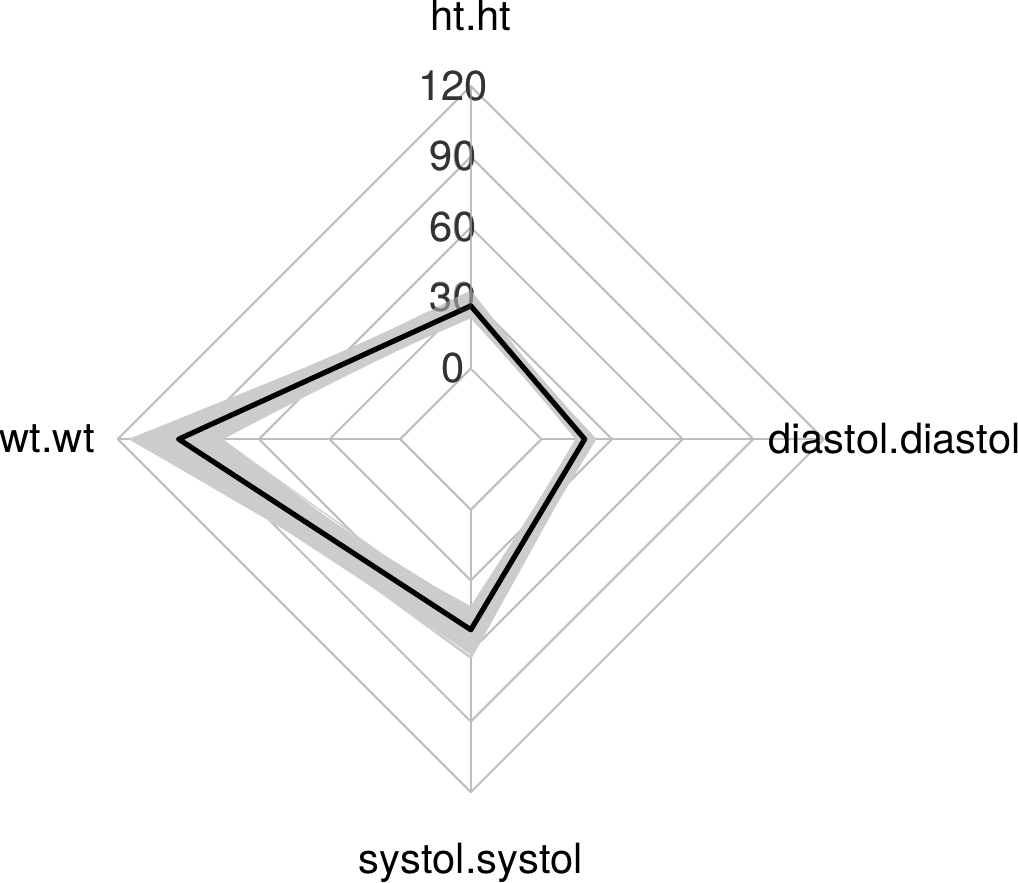}
\end{subfigure}
\begin{subfigure}{0.49\linewidth}
  \centering
  \includegraphics[scale=0.65]{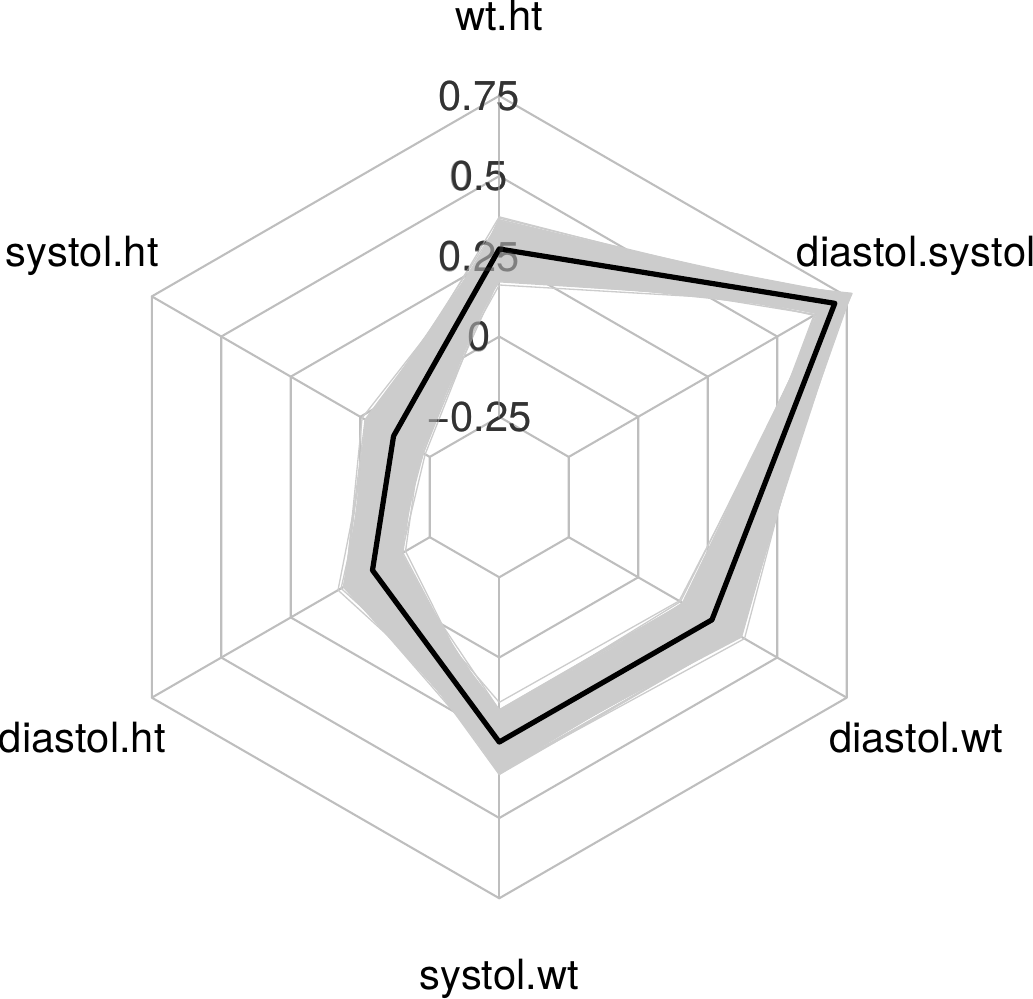}
\end{subfigure}
\caption{Radarcharts of the intrinsic sample mean (black) covariance matrix across clinical centers and 95-$\%$ intrinsic zonoid depth bootstrapped confidence region (grey) based on $20\, 000$ bootstrapped samples, equivalent to Figure \ref{fig:8} in the main document for the GDD. Again, we break ties in the center-to-outward zonoid depth-ranks by assigning the lowest rank to the clinical center with smallest Riemannian distance to the intrinsic sample mean.}
\end{figure}

\end{document}